\RequirePackage{fix-cm}
\documentclass[twocolumn]{svjour3}          
\smartqed  
\usepackage{graphicx}
\usepackage{lscape}
\usepackage{multicol}
\usepackage{sidecap}
\usepackage{caption}
\usepackage{soul}
\usepackage{amsmath,amssymb,amsfonts,latexsym,color,url}
\usepackage[authoryear]{natbib}

\def\sN{{\mathcal N}}

\def\sB{{\mathcal B}}
\def\sD{{\mathcal D}}

\def\sF{{\mathcal F}}

\def\vectorfontone{\bf}
\def\vectorfonttwo{\boldsymbol}
\def\vw{{\vectorfontone w}}                      %
\def\vx{{\vectorfontone x}}                      
\def\vy{{\vectorfontone y}}                      

\def\vone{{\vectorfontone 1}}

\def\vbeta{{\vectorfonttwo \beta}}               
\def\vgamma{{\vectorfonttwo \gamma}}             %
\def\vtheta{{\vectorfonttwo \theta}}             
\def\vlambda{{\vectorfonttwo \lambda}}           
\def\vmu{{\vectorfonttwo \mu}}                   
\def\vsigma{{\vectorfonttwo \sigma}}             %
\def\vphi{{\vectorfonttwo \phi}}                 %
\def\matrixfontone{\bf}
\def\matrixfonttwo{\boldsymbol}
\def\mI{{\matrixfontone I}}                      
\def\mW{{\matrixfontone W}}                      
\def\mX{{\matrixfontone X}}                      
\def\mSigma{{\matrixfonttwo \Sigma}}             %
\def\bE{{\mathbb E}}                             
\def\bP{{\mathbb P}}                             
\def\bR{{\mathbb R}}                             

\def\sRCVB{{\textnormal{\tiny\textsc{RCVB}}}}
\def\expit{\text{expit}}
\def\sJ{\mathcal{J}}

\def\ds{\displaystyle}

\newcommand{\indep}{\raisebox{0.05em}{\rotatebox[origin=c]{90}{$\models$}}}
\def\argmax{\operatornamewithlimits{\text{argmax}}}

\definecolor{darkgreen}{rgb}{0.0, 0.4, 0.1}

\begin{document}

\title{Variational discriminant analysis with variable selection
}

\titlerunning{Variational discriminant analysis with variable selection}        

\author{Weichang Yu         \and
        John T. Ormerod       \and
        Michael Stewart
}

\institute{Weichang Yu \at
              1. School of Mathematics and Statistics, University of Sydney, New South Wales, Australia \\
              \email{W.Yu@maths.usyd.edu.au}           \\
           \and
           John T. Ormerod \at
            1. School of Mathematics and Statistics, University of Sydney, New South Wales, Australia \\
            2. ARC Centre of Excellence for Mathematical \& Statistical Frontiers \\
            \email{john.ormerod@sydney.edu.au} \\
            \and
            Michael Stewart \at
            1. School of Mathematics and Statistics, University of Sydney, New South Wales, Australia \\
            \email{michael.stewart@sydney.edu.au}    
}

\date{Received: date / Accepted: date}

\maketitle

\begin{abstract}

A fast Bayesian method that seamlessly fuses classification and hypothesis testing via
discriminant analysis is developed. 
Building upon the original discriminant analysis classifier, modelling components are added to 
identify discriminative variables. A combination of cake priors and a novel form of variational
Bayes we call reverse collapsed variational Bayes gives rise to variable selection that can be 
directly posed as a multiple hypothesis testing approach using likelihood ratio statistics. Some 
theoretical arguments are presented showing that Chernoff-consistency (asymptotically zero type I 
and type II error) is maintained across all hypotheses. We apply our method on some publicly 
available genomics datasets and show that our method performs well in practice for
its computational cost. An {\tt R} package 
\texttt{VaDA} has also been made available on Github.

\keywords{Discriminant analysis \and Variational Bayes approximation \and Variable selection \and 
	Cake priors \and Multiple hypothesis tests \and Classification \and Fast algorithms}

\end{abstract}

\section{Introduction and literature review}
\label{intro}
Classification is a fundamental component of machine learning that is applicable in many 
disciplines. A popular classification method, initially known as discriminant analysis, was first 
introduced by \cite{Fisher1936} and has more recently been adapted by \cite{Dudoit2002} and 
\cite{Fernandez-Delgado} to achieve consistently good performance for some high dimensional 
datasets. This class of methods involves a comparison of group proportions and group-conditional 
distributions of variables, also known as features in machine learning literature, to arrive at a 
classification decision rule. However, 
this decision rule can not 
be computed when applying discriminant analysis (DA) to high dimensional data, i.e., when the number 
of observations, $n$, is less than number of variables, $p$. Furthermore, the standard DA
model of \cite{Fisher1936} is not designed to identify discriminative (or signal) variables. Without 
modification of the base DA model, its usage in high dimensional problems where signal
identification is important is limited, e.g., bioinformatics. Moreover, the gradual accumulation
of estimation errors as the number of noise variables increases can lead to a substantial loss in classification accuracy \citep{FanFan}.

In discriminant analysis, the group-conditional distribution of variables are 
commonly assumed to be Gaussian. 
This simplifies the classification rule to a difference in Mahalanobis distances 
of a new observation from the group-conditional distributions. 
Since the MLE of the covariance matrix required to compute the Mahalanobis distance is 
singular in high dimensional data, i.e., when $n < p$, the classification decision rule cannot be computed. 
A straightforward solution to this problem is to use alternative estimators such as 
the Moore-Penrose inverse \citep{Courrieu2005, Chen2014, CaiLiu} or an
alternative covariance matrix. For example, \cite{Thomaz2006} proposed a stabilised
covariance matrix, whereas \cite{Fisher2011} and \cite{Guo2007} used different forms of penalised estimators which
we shall describe in further details in Section \ref{Numericals}.

Another common solution is to utilise dimension reduction techniques such 
as principal components or t-distributed stochastic neighbour embeddings 
(t-SNEs)   \citep{vanderMaaten2017} to project the original variables into a lower dimensional space. 
The high dimensionality issue has also been tackled by making the \emph{na\"{i}ve 
Bayes} assumption, i.e., the covariance matrix for the variables is assumed to be diagonal
\citep{Dudoit2002}. 
Numerous examples of discriminant analysis models, sometimes called na\"{i}ve Bayes classifiers, 
that have made this 
assumption can be found in \cite{TibshiraniNSC, FanFan, Witten, WittenTibs}.
While dimension reduction solutions are simple to implement, they do not necessarily 
address the need to 
identify discriminative variables commonly required in high dimensional data
analysis.

An approach that addresses high dimensionality and identifies signal variables is to 
implement  a two-stage algorithm. In the first stage, a hypothesis test is performed on each of 
the variables to identify signal variables. In the second stage, variables that are 
identified as signals are retained and used to fit a DA model \citep[e.g.,][]{FanFan}. 
Care must be taken when choosing an appropriate variable selection method as such methods can lead to 
inflated family-wise type I error, false discovery rates or other multiple testing issues 
\citep{Shaffer1995}. This can be easily resolved with one of numerous remedial measures   
\citep[see for example,][]{Bonferroni1936, Benjamini1995, Benjamini2001, Storey2003}. 
A notable criterion, known as higher criticism thresholding, exhibits asymptotic optimality and 
good performance in several multiple testing metrics such as false discovery rate and missed 
detection rate under some sparsity assumptions \citep{DonohoJin2004, DonohoJin2008}. 
This criterion has been incorporated as an option in two-stage DA algorithms such as shrinkage DA 
({\tt R} package: {\tt SDA}) and factor-adjusted discriminant analysis ({\tt R} package: 
{\tt FADA}) \citep{Ahdesmaki, Perthame2016}. A modified version of the criterion, known as expanded 
higher criticism (EHC) \citep{Pedro}, has been incorporated into diagonal linear discriminant analysis (DLDA) and 
factor-based linear discriminant analysis (R package: \texttt{HiDimDA}). Although many of these 
algorithms utilise selection criteria that have good theoretical properties, information from the 
variable selection stage is lost when the variable selection and classification are done in two 
separate stages. For example, two variables yielding adjusted p-values of $0.001$ and $0.04$ may be 
selected, but their difference in signal strengths remain unaccounted for. This, in turn, may lead 
to an unnecessary loss of classification accuracy.

The loss of information can be circumvented by fusing the two stages. This fusion can be 
realised in penalised discriminant analysis models. 
In such methods a penalty function induces sparsity in the estimated discriminant vector (product 
of precision matrix and mean difference), that is used for both variable selection and 
classification. \cite{WittenTibs} introduced two penalty options to the Fisher's discriminant 
problem ({\tt R} package: {\tt penalizedLDA}) which will be further elaborated in the Section 
\ref{Numericals}, whereas an $L_1$ penalty is introduced through a regression framework
in \cite{Clemmensen2011} and \cite{Mai2012}. Other examples may be found in \cite{CaiLiu}, \cite{Shao} and \cite{SafoAhn}. 
While penalised DA models have demonstrated desirable theoretical properties and good numerical 
results in these papers, the results are often very sensitive to the setting of the tuning parameter 
of the penalty function. Usually costly cross-validation is often necessary to determine an 
appropriate value of this tuning parameter. 

In this paper, we propose a   Bayesian DA model 
that integrates both variable selection and classification. 
The model overcomes high dimensionality by adopting the na\"{i}ve Bayes assumption leading to an 
invertible estimated covariance matrix. There are numerous works in the literature
that either criticise or justify this assumption.
The loss in classification accuracy under this assumption was demonstrated through 
several simulation studies especially when the correlation structure is complex \citep{Clemmensen2013, Perthame2016} and
theoretical criticisms were made in \cite{Mai2012}. However, na{\"i}ve Bayes
DA models demonstrated excellent classification performance in several studies on
publicly available gene expression datasets \citep{Dudoit2002, TibshiraniNSC, FanFan, Witten}
and its worst-case misclassification error has good asymptotic convergence properties
under appropriate conditions on the covariance matrix \citep{Bickel}. The substantially lower computational cost associated with
a diagonal covariance matrix also makes the assumption attractive. 
In our view, the key issue here is a potential trade-off between faster computational speed obtained
by using the na{\"i}ve Bayes assumption and improvement in classification accuracy by accounting for correlation. We believe when
there is low to moderate correlation between predictors, or for very high dimensional
problems that this trade-off favours 
na{\"i}ve Bayes DA.

Unlike the two-stage methods, the Bayesian hierarchical setup of our proposed model fuses both variable 
selection and classification in an omnibus fashion. By doing so we avoided any loss of information 
between the variable selection and classification stages. By introducing variable selection parameters into our Bayesian model, we allow posterior 
	inferences to be made on the \emph{discriminativeness} of each variable. The variable selection component fits 
naturally into a multiple hypothesis testing (MHT) paradigm. By choosing the cake priors 
\citep{OrmerodCake} and approximating the computationally intensive posterior densities with a new 
variant of variational Bayes inference known as reverse collapsed variational Bayes (RCVB), 
the resultant decision rule overcomes the problems associated with MHTs such as inflated type I 
errors. Since the variable selection rule depends on a set of approximate posterior probabilities, 
the choice of the selection threshold can also be intuitively determined. The resultant 
classification rule takes the form of a weighted na\"{i}ve Bayes linear discriminant analysis
(when variances are assumed equal). The computational cost of the algorithm is also reduced by 
utilising Taylor's approximation to reduce the number of updates in the RCVB cycles. An 
implementation of our approach uses \texttt{C++} for high performance computing. These endeavours
to keep the computational cost low are aimed at ensuring the scalability of our models
to many high dimensional gene expression datasets when computing resources are limited.

In Section 2 we specify the model and discuss our choice of priors. 
Section 3 introduces our RCVB approximation. 
We will discuss its application to our model in Section 4. 
In Section 5, we state some asymptotic properties of our variable selection 
criterion induced by our proposed model and hence show that our variable 
selection rule circumvents issues with MHTs. 
In Section 6 
we compare the performance of our proposed model 
with existing solutions by an application to simulated and publicly available datasets and 
discuss the choice between two versions of our proposed model. Section 7 concludes.

\section{The variational discriminant analysis model}
\label{Model}

Consider the training dataset $\{ \vx_i, y_i \}_{i=1}^{n}$ where for each $i$
we have  $\vx_i = (x_{i1},\ldots,x_{ip})^T$ as a vector of predictor-variables 
and $y_i \in \lbrace 0, 1 \rbrace$ as an observed group label. 
In this paper, we present the model in the context of a binary classification problem 
but its extension to multiple groups should be possible. 
In addition, we have used bold-faced symbols to denote the vector of parameters 
across subscripts. 
We assume that $y_i$ is observed for all $i = 1, \ldots, n$. 
In line with machine learning terminology, 
we shall refer to this dataset with observed group labels as the training data.

We consider the following hierarchical model for our data set. 
For all $i=1,\ldots,n$, each $(\vx_i, y_i)$ is distributed as follows. 
Let $\gamma_j\in\{0,1\}$, $1\le j\le p$ be a binary variable indicating whether
variable $j$ is discriminative. If $\gamma_j=1$,
then\\
\begin{equation}\label{eq:alternative}
\hspace{3mm} \left\{ \begin{array}{ll}
x_{ij} \;| \; y_i, \mu_{j1},\sigma_{j 1}^2 \stackrel{\mbox{\scriptsize iid}}{\sim} 
\sN(\mu_{j 1}, \sigma_{j 1}^2) & \mbox{ if $y_i = 1$; \;} \\ [1ex]
x_{ij} \; |\; y_i, \mu_{j0},\sigma_{j 1}^2 \stackrel{\mbox{\scriptsize iid}}{\sim} 
\sN(\mu_{j 0}, \sigma_{j1}^2) & \mbox{ if $y_i = 0$},
\end{array} \right.
\end{equation}

\noindent and if $\gamma_j = 0$, then
\begin{equation}\label{eq:null}
x_{ij} \, | \, \mu_{j},\sigma_{j}^2 \, \stackrel{\mbox{\scriptsize iid}}{\sim} \, \sN(\mu_{j }, \sigma_{j}^2),
\end{equation}

\noindent where $\sN(m, s^2)$ denotes the Gaussian distribution with mean $m$ and variance $s^2$. When $\gamma_j=1$, discriminativeness is induced by imposing dependence 
between the Gaussian parameters and group label $y_i$. We model the conditional distribution of the group labels as
$$
y_i \, | \, \rho_{y} \,
\stackrel{\mbox{\scriptsize iid}}{\sim} \, \text{Bernoulli}(\rho_y),
$$

\noindent 
where $\rho_y$ denotes the probability of observing a group 1 sample from the population.

To avoid the computational problem of inverting huge covariance matrices, we make the
the na\"{i}ve Bayes assumption by imposing conditional independence 
between the variables given the Gaussian parameters 
and group labels, i.e. $x_{ij} \, \indep \, x_{ik}$ for any $j \neq k$.

\subsection{Homogeneity of group-specific variances}
\label{VLDAVQDA}
We have assumed that the group conditional variances are equal, 
i.e,. Var$(x_{ij} | y_i = 1, \sigma_{j 1}^2) = $ Var$(x_{ij} | y_i = 0, \sigma_{j 1}^2)$. 
Gaussian discriminant analysis models that adopt this assumption are 
known as linear discriminant analysis (LDA) since the resultant decision 
rule is linear in the variables $\vx$ \citep[see][]{Bickel, Ahdesmaki, Pedro, WittenTibs, Clemmensen2011, Perthame2016}.  
Those that allow for different group conditional variances are known 
as quadratic discriminant analysis (QDA) \citep{Srivastava2007}.  The choice between LDA
and QDA boils down to conditions which LDA is robust to the violation of variance homogeneity.
Some simulation studies \citep{Marks1974, Zavorka2014} led to the unexpected
conclusion that LDA exhibits better performance than QDA when the sample size is small and the 
departure from the variance homogeneity assumption is mild. As pointed out by a reviewer, this superiority can be understood
by observing that LDA has lesser parameters than QDA. Consequently, the 
more efficient parameter estimation compensates for the model misspecification.
In spite of existing findings, 
superior performance by LDA cannot be guaranteed in publicly available datasets
where it is plausible for us to have many true signals with large differences in group-conditional variances.
While a compromising solution between QDA and LDA has been proposed by \cite{Friedman1989}, 
we propose two variants of our model - a variational linear 
discriminant analysis (\texttt{VLDA}) and a variational quadratic 
discriminant analysis (\texttt{VQDA}), 
that correspond to LDA and QDA methods respectively. 
This
allows us to study the robustness of na{\"i}ve Bayes LDA under varying differences in the group-conditional variances
and have a QDA version of our proposed model ready when LDA fails.

\subsection{Choice of priors}
\label{Priors}

The choice of priors is of paramount importance in Bayesian modelling as it affects the computational 
course of the posterior inference and allows prior knowledge to influence results of the analysis. 
A natural choice of priors for $\vgamma$ and $\rho_y$ uses
$$
\gamma_j \; | \; \rho_{\gamma} \stackrel{\mbox{\scriptsize iid}}{\sim} \text{Bernoulli} (\rho_\gamma), \quad \mbox{with} \quad 
\rho_\gamma \sim \text{Beta} (a_\gamma, b_\gamma),
$$

\noindent
and
\begin{align*}
&\rho_y \sim \text{Beta} (a_y, b_y),
\end{align*}
where the random parameter $\rho_{\gamma}$ may be interpreted as the probability of including 
a true signal in the dataset. 

For the rest of this paper, we have chosen a flat prior for $\rho_y$ by assigning the hyperparameters
$a_y = b_y = 1$. 
When considering settings for $a_\gamma$ and $b_\gamma$, we assume that the 
true model consists of $p_0$ noise and $p_1$ signal variables, and that
the total number of variables grow at a non-polynomial rate of $n$. Here, we propose 
the setting $a_\gamma = 1$ and 
\begin{equation}\label{eq:b_gamma} 
\ds b_\gamma = \frac{p_n^2}{\sqrt{n+1}} \exp\left[ \frac{\kappa (n+1)}{ \log(n+1)^{r}} \right] 
\end{equation} 

\noindent for some $r<1$ and $\kappa > 0$, 
to induce desirable asymptotic 
properties in our resultant variable selection rule (see Section \ref{Asymptotics}).

The choice of priors for the Gaussian parameters is a more complex issue. Most research is 
conducted in the absence of reliable prior knowledge and therefore diffuse priors are a popular 
choice. However, a diffuse prior may not retain the ``diffuse" property after a model 
reparametrisation. Besides this property, we also require our priors to lead to model selection 
consistency in our variable selection rule and strong concordance with frequentist approaches 
(information consistency) in both variable selection and classification rules. The latter criterion 
would help avoid unnecessary dilemma when users compare the inference results of our model with most 
other discriminant analysis models that are frequentist in approach.

Recently, \cite{OrmerodCake} has proposed the cake prior which can be made diffuse under some 
setting of hyperparameters. For a two-sample test of Gaussian data, they have also obtained 
Bayes factors which have good asymptotic properties. Furthermore, the Bayes 
factor takes the form of a penalized likelihood ratio statistic. The cake priors for 
our Gaussian parameters are as follows.
For \texttt{VLDA} the model under $H_{0j}$ for variable
$j$ is
(\ref{eq:null}) and the model under $H_{1j}$ for variable
$j$ is (\ref{eq:alternative}). The cake priors for these
hypotheses are given by
\begin{align*}
&\mu_{j} | \sigma_{j}^2 \sim \sN(0, h^{1/2} \sigma_{j}^2), \;\;\; \sigma_{j}^2 \stackrel{\mbox{\scriptsize iid}}{\sim} LN(0, 2h^{1/2}), \\
&\mu_{jk} | \sigma_{j1}^2 \sim \sN(0,  (n / n_k) h^{1/3} \sigma_{j1}^2)  \;  \text{for} \; k =0,1, \\
&\text{and } \sigma_{j1}^2 \stackrel{\mbox{\scriptsize iid}}{\sim} LN(0, 2 h^{1/3} ).
\end{align*}

\noindent 
where $n_1 = \sum_{i=1}^{n} y_i$, $n_0 = n - n_1$, $LN$ refers to the log-normal distribution 
and $h>0$ is a common hyperparameter shared between $\{H_{0j}\}_{j=1}^p$
and $\{H_{1j}\}_{j=1}^p$.

Analogously, the \texttt{VQDA} model under $H_{1j}$ is
\begin{equation}\label{eq:alternativeQDA}
\hspace{3mm} \left\{ \begin{array}{ll}
x_{ij} \;| \; y_i, \mu_{j1},\sigma_{j 1}^2 \stackrel{\mbox{\scriptsize iid}}{\sim} 
\sN(\mu_{j 1}, \sigma_{j 1}^2) & \mbox{ if $y_i = 1$; \;} \\ [1ex]
x_{ij} \; |\; y_i, \mu_{j0},\sigma_{j 0}^2 \stackrel{\mbox{\scriptsize iid}}{\sim} 
\sN(\mu_{j 0}, \sigma_{j 0}^2) & \mbox{ if $y_i = 0$},
\end{array} \right.
\end{equation}
and is equivalent to (\ref{eq:null}) under $H_{0j}$ so that the group conditional variances differ. The cake priors in this case are given by
\begin{align*}
&\mu_{j} | \sigma_{j}^2 \sim \sN(0, h^{1/2} \sigma_{j}^2), \;\;\; \sigma_{j}^2 \stackrel{\mbox{\scriptsize iid}}{\sim} LN(0, 2h^{1/2}), \\
&\mu_{jk} | \sigma_{jk}^2 \sim \sN(0,  (n / n_k) h^{1/4} \sigma_{jk}^2 ), \; \text{ for } \; k =0,1, \\
&\text{and } \sigma_{jk}^2 \stackrel{\mbox{\scriptsize iid}}{\sim} LN(0, 2 (n / n_k) h^{1/4}  ), \; \text{ for } \; k =0,1,
\end{align*}

\noindent 
 The interested reader may refer to \cite{OrmerodCake} for details about the 
cake prior construction.  
Although cake priors do not lead to closed form expressions 
for the posterior of $\vgamma$ and the group labels of
new observations, we will observe in Section \ref{Inference} that they yield approximate posteriors that 
satisfy both model selection and information consistency when implemented with a posterior 
inference method described in the next section.
\section{Approximating the posterior}
\label{ApproxPosterior}

For most models posterior distributions are known only up to a normalising constant. While  Markov chain Monte Carlo (MCMC) 
methods are the most widely used methods for making inferences from such posteriors, a new class of fast, 
deterministic algorithms known as variational Bayes approximation is gaining popularity in the computer 
science literature and has demonstrated comparable results at a fraction of MCMC's computational cost 
in complex problems presented by \cite{bleiDPmixture}, and \cite{Luts2014}. 

We now provide a review of variational Bayes (VB) approximation and then describe a modification of this method 
called reverse collapsed variational Bayes. For a more comprehensive
introduction to VB the reader 
may refer to \cite{ormerodBasic} and \cite{Blei2017}.

\subsection{Variational Bayes}
\label{VB}
Given a model's parameter $\vtheta$ and data $\sD$, the posterior density 
of $\vtheta$ may be expressed as
\begin{align*}
p(\vtheta | \sD) &= \frac{p(\vtheta, \sD)}{ p(\sD) }.
\end{align*}
The denominator is known as the \emph{marginal likelihood} of the data and involves the evaluation of 
an integral or a sum that may be computationally infeasible. The endpoint of the variational Bayes (VB) 
algorithm is to choose an approximation $q(\vtheta)$ to the posterior density 
$p(\vtheta | \sD)$ from a set of functions $\sF$ that are more computationally feasible 
by minimising the Kullback-Leibler (KL) divergence
\begin{align} \label{eq:1}
D_{KL} (q || p ) &= \mathbb{E}_{q} \bigg [ \log \bigg \lbrace 
\frac{ q(\vtheta)}{p(\vtheta | \sD) }  \bigg \rbrace \bigg ],
\end{align}
where $\mathbb{E}_{q}$ refers to the expectation with respect to $q(\vtheta)$. Since
\begin{align*}
&\bE_{q} \bigg [ \log \bigg \lbrace \frac{ q(\vtheta)}{p(\vtheta | \sD) }  \bigg \rbrace \bigg ] 
= \bE_{q} \bigg [ \log \bigg \{ \frac{ q(\vtheta) p(\sD)} {p(\vtheta , \sD) } \bigg \} \bigg ], \\
& \; \; \; \; \; \; \; \; \;= \log p(\sD) - \mathbb{E}_{q} \bigg [ \log \bigg \{ \frac {p(\vtheta , \sD) } { q(\vtheta)} \bigg \} \bigg ],
\end{align*}
minimising (\ref{eq:1}) is also equivalent to maximising the Expected Lower Bound Order (ELBO) given by
\begin{equation} \label{SVBELBO}
\mbox{ELBO}_{\text{VB}} = \bE_{q} \bigg [ \log \bigg \{ \frac {p(\vtheta , \sD) } { q(\vtheta)} \bigg \} \bigg ].
\end{equation}
\noindent One common choice for $\sF$ is the set of mean field functions $\sF = \lbrace q(\vtheta) \; | \; q(\vtheta) 
= \prod_{j=1}^{J} q_j(\vtheta_j) \rbrace$, where $\{ \vtheta_j \}_{j=1}^J$ is a partition of
$\vtheta$. This choice of $\sF$ leads to the optimal approximating densities
\begin{align} \label{VBqdensity}
{q}_j (\vtheta_{j}) \propto &\exp \left [ \bE_{-{q}_j} \big \lbrace \log p( \vtheta, \sD ) \big \rbrace \right ], \, \forall \; j=1,\ldots,J,
\end{align}
as shown, for example, in \cite{ormerodBasic}. The notation $\mathbb{E}_{-{q}_j}$ refers to expectation with respect to $\prod_{l \neq j} q_{l} (\vtheta_{l})$. 
The parameters of each ${q}_j (\vtheta_{j})$ is updated iteratively with the batch coordinate-ascent variational inference algorithm as described 
in \cite{Zhang2017}.

 
\subsection{Reverse collapsed variational Bayes}

\label{CVB}
A variant of VB called collapsed variational Bayes (CVB) was first coined in the context of latent 
Dirichlet allocation \citep{tehLDA}. The key idea behind these methods is to collapse, or 
marginalise over a subset of parameters before applying VB methodology. The CVB approach of 
\cite{tehLDA} results in a better approximation in comparison to VB, but is no different 
conceptually since it can be thought of simply as applying VB to the marginalized likelihood. We now 
introduce a \emph{reverse collapsed variational Bayes} (RCVB) which can result in a different
approximation to CVB. In this form the lower bound is calculated by using VB for one set of 
parameters, and the remaining set of parameters are collapsed over by marginalization.

To fix ideas, let $\vtheta_1$ and $\vtheta_2$ be a partition of the parameter vector $\vtheta$. 
Suppose we have a density $q_2(\vtheta_2)$ such that the quantity
$$
\log \underline{p}(\sD,\vtheta_1) \equiv \bE_{q_2(\vtheta_2)} \log\bigg\{ \frac{p(\sD,\vtheta_1,\vtheta_2)}{q_2(\vtheta_2)}\bigg\}
$$

\noindent can be evaluated analytically for all $\vtheta_1$. Using
Jensen's inequality it is easy to show that 
$$
\log p(\sD,\vtheta_1) \ge \log \underline{p}(\sD,\vtheta_1)
$$

\noindent for all $\vtheta_1$. If we use 
$\log \underline{p}(\sD,\vtheta_1)$ as an approximation of 
$\log p(\sD,\vtheta_1)$ we can then marginalise over $\vtheta_1$ 
to obtain the following lower bound on the log marginal likelihood:
$$
\begin{array}{rl}
\mbox{ELBO}_{\text{RCVB}} 
& \ds \equiv \log \int  \underline{p}(\sD,\vtheta_1) d\vtheta_1, \\
& \ds =\log \int \exp\bigg[ \bE_{q_2} \log\bigg\{ \frac{p(\sD,\vtheta_1,\vtheta_2)}{q_2(\vtheta_2)}\bigg\} \bigg] d\vtheta_1,
\end{array} 
$$

\noindent 
where the integral can be interchanged with a sum when appropriate. Since
$\log p(\sD,\vtheta_1) \ge \log \underline{p}(\sD,\vtheta_1)$ we have
$\mbox{ELBO}_{\text{RCVB}}  \ge 
\mbox{ELBO}_{\text{VB}}$ and hence RCVB yields an approximation of
the log marginal likelihood
that is more accurate than VB.

Now suppose that we have partitioned $\vtheta$ into three sets of parameters $\vtheta_1$, 
$\vtheta_2$ and $\vtheta_3$, and we want to apply VB-type approximations
to $\vtheta_2$ and $\vtheta_3$ while integrating out $\vtheta_1$
analytically. The iterations for RCVB algorithms
would be to repeat the following two steps for an arbitrarily large number
of iterations until convergence:
\begin{equation}\label{RCVBupdates}
\begin{array}{rl}
1. & \ds q_2(\vtheta_2) \propto \int \exp\left[ \bE_{q_3} 
\log\left\{ \frac{p(\sD,\vtheta_1,\vtheta_2,\vtheta_3)}{q_3(\vtheta_3)}\right\}  \right] d\vtheta_1.  \\
& \\
2. & \ds q_3(\vtheta_3) \propto \int \exp\left[ \bE_{q_2} 
\log\left\{ \frac{p(\sD,\vtheta_1,\vtheta_2,\vtheta_3)}{q_2(\vtheta_2)}\right\}  \right] d\vtheta_1.
\end{array}  
\end{equation}

Similarly to VB, the parameters of each ${q}_j (\vtheta_{j})$ is updated iteratively with the 
batch coordinate-ascent variational inference algorithm.

The above ideas are easily generalizable to an arbitrary partition size.
In the next section we demonstrate the use of RCVB to perform posterior inference for our proposed model.

\section{Posterior inference in variational discriminant analysis}
\label{Inference}
The endpoint of variational discriminant analysis with variable selection (VaDA) 
is to identify the signal variables and the true group label of new observations at a  low computational cost. 
The calculations provided in the rest of this section will pertain only to \texttt{VLDA}. 
Posterior inference calculations for \texttt{VQDA} are provided in the appendix. 
Given $m$ new observations $\{(\vx_{n+i}, y_{n+i})\}_{i=1}^{m}$ where 
$\{y_{n+i}\}_{i=1}^m$ are latent, we may regard $ \boldsymbol{\gamma}$ and 
$\{y_{n+i}\}_{i=1}^m$ as the parameters of interest. Let
$$\vtheta_1 = (\vmu_1, \vmu_0, \vmu, \vsigma_1^2, \vsigma^2, \rho_y, \rho_\gamma)$$ be the 
parameters not required for inference, which will be collapsed over. We provide details about the inference for 
$m=1$ and will describe the generalisation to any $m>1$ towards the end. 
Let $\mX = [\vx_1, \ldots, \vx_{n}]^T = [\widetilde{\vx}_1, \ldots, \widetilde{\vx}_p] \in \bR^{n\times p}$ , 
where $\widetilde{\vx}_j$ is the column vector of variable $j$ of the data and 
$\vy = (y_1, \ldots, y_{n})^T$ is the observed column vector of binary responses.
Let $\sD = (\mX, \vy)$ denote the training data.  
The posterior distribution of $\gamma_j$ given $\sD$ and $\vx_{n+1}$ may be expressed as
\begin{equation*}
\ds p(\gamma_j \; | \; \sD, \vx_{n+1}) = \frac{ p(\gamma_j, \sD, \vx_{n+1})}{ p(\sD, \vx_{n+1}) }.
\end{equation*}

Following arguments for a diffused prior in \cite{OrmerodCake}, we let 
$h \rightarrow \infty$. The marginal likelihood of the observed data ($\sD$ and $\vx_{n+1}$) in the denominator 
can be written as
\begin{equation} \label{margLik}
\begin{array}{l} 
p(\sD, \vx_{n+1})  \\ 
\ds = \sum_{{\tiny y_{n+1} \in \{ 0, 1 \}}} \sum_{\vgamma \in \{ 0, 1 \}^p} 
\int p(\sD, \vx_{n+1}, y_{n+1},\vgamma, \vtheta_1) \; \text{d} \vtheta_1  \\
\ds = \left[ \frac{\prod_{j=1}^p p(\widetilde{\vx}_j, x_{n+1,j}| \gamma_j = 0)}{\sB(a_y,b_y) 
	\sB(a_\gamma,b_\gamma)} \right]  \\ 
\ds \times \sum_{y_{n+1} \in \{ 0, 1 \}} 
\Bigg [ \sB(a_y + n_1 + y_{n+1}, b_y + n_0 + 1 - y_{n+1}) \\ 
\ds \times \sum_{\vgamma\in\{0,1\}^p}  \exp\Big\{ \log \sB(a_\gamma 
+ \vone^T\vgamma, b_\gamma + p - \vone^T\vgamma)  \\ 
\ds \hspace{7mm} + \tfrac{1}{2} \vgamma^T\vlambda_{\mbox{\scriptsize Bayes}} (\widetilde{\vx}_j, x_{n+1,j}, \vy, y_{n+1}) \Big\} \Bigg],
\end{array} 
\end{equation} 

\noindent 
where $\sB(a,b) = \Gamma(a)\Gamma(b) / \Gamma(a+b)$ is the beta function, $n_1 = \vone^T \vy$, $n_0 = n - n_1$,
and $\vlambda_{\mbox{\scriptsize Bayes}}$ is a column vector of size $p$ which is defined as follows. 
The likelihood ratio statistic corresponding to 
the test
$\{H_{0j}: \gamma_j = 0 \}$ using model
(\ref{eq:null}) against $\{H_{1j}: \gamma_j = 1 \}$ which uses (\ref{eq:alternative}) for  variable $j$ is
\begin{equation*}
\lambda_{\text{LRT}} (\widetilde{\vx}_j, x_{n+1,j},  \vy, y_{n+1}) = (n+1) \log 
\left ( \frac{\widehat{\sigma}_{j}^2}{\widehat{\sigma}_{j 1}^2} \right ),
\end{equation*}

\noindent where the maximum likelihood estimates (MLEs) are
\begin{align*}
&\widehat{\sigma}_{j 1}^2 = \frac{1}{n+1} \bigg [ ||\vy^T\{\widetilde{\vx}_{j} 
- \widehat{\mu}_{j1}\vone\}||^2 \\ & \hspace{17mm} + ||(\vone - \vy^T)\{\widetilde{\vx}_{j} - \widehat{\mu}_{j0}\vone\}||^2 \\
&\hspace{3mm} + y_{n+1} (x_{n+1,j} - \widehat{\mu}_{j1})^2 
+ (1 - y_{n+1}) (x_{n+1,j} - \widehat{\mu}_{j0})^2 \bigg ], \\
&\widehat{\sigma}_{j}^2 =  \tfrac{1}{n+1} \left \{  ||\widetilde{\vx}_{j} 
- \widehat{\mu}_{j0}\vone||^2 + (x_{n+1,j} - \widehat{\mu}_{j})^2 \right \}, \\
&\widehat{\mu}_{j 1} = \tfrac{1}{n_1+y_{n+1}} \Big \{\vy^T \widetilde{\vx}_j + y_{n+1 }x_{n+1,j} \Big \},\\ 
&\widehat{\mu}_{j 0} = \tfrac{1}{n_0+1 - y_{n+1}} \Big \{ (\vone - \vy)^T \widetilde{\vx}_j + (1-y_{n+1 })x_{n+1,j} \Big \}\\
&\widehat{\mu}_{j} = \tfrac{1}{n+1} \Big \{\vone^T \widetilde{\vx}_j + x_{n+1,j} \Big \}.
\end{align*}

\noindent Then
the $j^{\text{th}}$ entry of $\vlambda_{\mbox{\scriptsize Bayes}}$ is (as $h\rightarrow \infty$)
\begin{align*}
\lambda_{\mbox{\scriptsize Bayes}} &(\widetilde{\vx}_j, x_{n+1,j}, \vy, y_{n+1}) \\ 
&\rightarrow \lambda_{\text{LRT}} (\widetilde{\vx}_j, x_{n+1,j}, \vy, y_{n+1}) - \log(n+1).
\end{align*}
\noindent The marginal likelihood in equation (\ref{margLik}) involves 
a combinatorial sum over $2^{p+1}$ binary combinations. Hence, exact 
Bayesian inference is computationally infeasible for large $p$ and  
approximation is required. We will use RCVB 
to approximate the posterior $p(\vgamma, y_{n+1} | \sD, \vx_{n+1})$ using the partition
\begin{align*}
q^{\sRCVB}(y_{n+1}, \boldsymbol{\gamma}) = q^{\sRCVB}(y_{n+1}) 
\prod_{j=1}^{p} q_j^{\sRCVB} (\gamma_j).
\end{align*}
We shall henceforth drop the superscript RCVB. Note that since 
$\gamma_j \in \{0,1\}$ and $y_{n+1} \in \{ 0,1\}$ we have 
\begin{align*}
&q_j (\gamma_j) = w_j^{\gamma_j} (1 - w_j)^{1 - \gamma_j},
\end{align*}
and
\begin{align*}
&q (y_{n+1}) = \tilde{y}^{y_{n+1}} (1 - \tilde{y})^{1 - y_{n+1}},
\end{align*}
where $\vw$ and $\widetilde{y}$ are \emph{variational parameters} to be optimized 
over, i.e., $q_j (\gamma_j)$ and $q(y_{n+1})$ are densities corresponding to a Bernoulli$(w_j)$ 
and Bernoulli$(\widetilde{y})$ distribution. 
The variational parameter $w_j$ may be interpreted as the approximate posterior probability 
that the hypothesis $H_{1j}$ is true when tested against $H_{0j}$. 
The interpretation of $\widetilde{y}$ is analogous. 

\subsection{Variable selection}
\label{VariableSelection}
With reference to the steps in (\ref{RCVBupdates}), we will apply VB approximation over 
$\vgamma_{-j}$ and $y_{n+1}$, and integrate analytically over $\vtheta_1$, i.e., for $1\leq j \leq p$ we have
\begin{align*}
&q_j(\gamma_j) \propto \int \exp \big [ \bE_{-q_j} \{ \log p(\sD, \vx_{n+1}, y_{n+1},\vgamma, \vtheta_1) \}  \big ] \text{d} \vtheta_1, \\
&\propto \exp\bigg[ \bE_{-q_j} \Big \{ \log\sB(a_\gamma + \vone^T\vgamma, b_\gamma + p - \vone^T\vgamma) \Big \} \\ 
&\hspace{12mm}+ \tfrac{\gamma _j}{2} \bE_{-q_j} 
\Big \{ \lambda_{\mbox{\scriptsize Bayes}} (\widetilde{\vx}_j, x_{n+1,j}, \vy, y_{n+1}) \Big \} \bigg ].
\end{align*}

\noindent For a sufficiently large $n$, we can avoid the need to evaluate the expectation 
$ \bE_{-q_j} \Big \{ \lambda_{\mbox{\scriptsize Bayes}} (\widetilde{\vx}_j, x_{n+1,j}, \vy, y_{n+1}) \Big \}$
by applying Taylor's expansion to approximate the MLEs with
\begin{align}
\label{TaylorExp}
&\bE_{-q_j} \log(a_\gamma + \vone^T \vgamma_{-j}) \approx \log(a_\gamma + \vone^T \vw_{-j}), \nonumber \\
&\bE_{-q_j} \log(b_\gamma + p - \vone^T \vgamma_{-j} - 1) \nonumber \\ 
&\hspace{8mm} \approx \log(b_\gamma + p - \vone^T \vw_{-j} - 1), \nonumber \\
&\widehat{\sigma}_{j 1}^2 \approx \tfrac{1}{n} \bigg [ ||\vy^T\{\widetilde{\vx}_{j} - \widehat{\mu}_{j1}\vone\}||^2 
+ ||(\vone - \vy^T)\{\widetilde{\vx}_{j} - \widehat{\mu}_{j0}\vone\}||^2 \bigg ], \nonumber \\
&\widehat{\sigma}_{j}^2 \approx  \tfrac{1}{n} ||\widetilde{\vx}_{j} - \widehat{\mu}_{j}\vone||^2, \nonumber \\
&\widehat{\mu}_{j 1} \approx \tfrac{1}{n_1} \vy^T \widetilde{\vx}_j , 
\;\; \widehat{\mu}_{j 0} \approx \tfrac{1}{n_0} (\vone - \vy)^T \widetilde{\vx}_j, 
\;\; \widehat{\mu}_{j} \approx \tfrac{1}{n} \vone^T \widetilde{\vx}_j,
\end{align}

\noindent and hence $\vlambda_{\mbox{\scriptsize Bayes}}$ does 
not depend on the new observation $(\vx_{n+1}, y_{n+1})$.
By using the approximation in (\ref{TaylorExp}), we have
\begin{align*}
&w_j = \frac{q_j(\gamma_j = 1)}{q_j(\gamma_j = 1) + q_j(\gamma_j = 0)}, \\
&\approx \expit\bigg[ \log (a_\gamma + \vone^T \vw_{-j}) - \log (b_\gamma + p - \vone^T \vw_{-j} - 1) \\
&\hspace{12mm}+ \tfrac{1}{2} \lambda_{\mbox{\scriptsize Bayes}} (\widetilde{\vx}_j, \vy)  \bigg ], \\
&= \expit\bigg[ \log (a_\gamma + \vone^T \vw_{-j}) - \log (b_\gamma + p - \vone^T \vw_{-j} - 1) \\
&\hspace{12mm}- \tfrac{1}{2} \log(n+1) + \tfrac{1}{2} \lambda_{\mbox{\scriptsize LRT}} (\widetilde{\vx}_j, \vy)   \bigg ], 
\\
&= \expit\bigg[ \mbox{penalty}_{j} + \tfrac{1}{2} \lambda_{\mbox{\scriptsize LRT}} (\widetilde{\vx}_j, \vy) \bigg ],
\end{align*}

Each $w_j$ may be viewed, from 
a frequentist perspective, as a ``test statistic" for $H_{1j}$ against $H_{0j}$. 
Hence, a natural decision rule is to identify variable $j$ as a signal if $w_j > c_w$
for some constant $c_w \in (0,1)$.

The $\mbox{penalty}_{j}$ term in the expression for $w_{j}$ can be interpreted as a data 
dependent penalty term which trades off the probability of type I errors against power. This presents
the task of identifying signal variables in our model as a multiple hypothesis
testing problem using penalized likelihood ratio statistics.
The constant $b_\gamma$ is particularly important. While we have chosen 
$a_\gamma=1$ and 
$b_\gamma$ given by (\ref{eq:b_gamma}) 
for some $r<1$ and $\kappa > 0$, one may specify other values of $b_\gamma$. If $b_\gamma$ 
is too small, then  false positives will occur when $p$ is allowed to diverge 
with $n$. If $b_\gamma$ is too large,
then there will be potentially too many 
false negatives. Ideally we want $b_\gamma$ 
to be as small as possible whilst 
having asymptotically zero false positives.
Since the updates of each $w_j$ no longer depends on $\widetilde{y}$ after applying the Taylor's expansion results, each
RCVB cycle only involves the update of $w_j$ over $1\le j\le p$ until convergence.

\subsection{Classification}
\label{Classification}
We will apply variational Bayes approximation over $\vgamma$ and integrate 
analytically over $\vtheta_1$ to obtain the approximate density for $y_{n+1}$, i.e.,
\begin{align*}
&q(y_{n+1}) \propto \int \exp \big [ \bE_{-y} \{ \log p(\sD, \vx_{n+1}, y_{n+1},\vgamma, \vtheta_1) \}  \big ] \text{d} \vtheta_1, \\
&\propto \exp\bigg[ \log\sB(a_y + n_1 + y_{n+1}, b_y + n_0 + 1 - y_{n+1}) \\ 
&\hspace{12mm}- \tfrac{1}{2} (n+1) \vw^{T} \log \Big \{(n+1)\widehat{\vsigma}_1^2 \Big \}  \bigg ],
\end{align*}

\newpage
\bgroup
\def\arraystretch{2.4}
\begin{table} \refstepcounter{table}\label{Algo1}
	\centering
	\begin{tabular}{l}
		\normalsize
		\bf{Table 1}~ \normalfont{Iterative scheme for obtaining the parameters in the optimal densities } $q\*(\boldsymbol{\gamma}, y_{n+1}, \ldots, y_{n+m})$ \normalfont{in \texttt{VLDA}}\\
		\hline
		\normalsize Require: For each $j$, initialise $w_j^{(0)}$ with a number in $[0,1]$.\\
		\normalsize \textbf{while} $||\vw^{(t)} - \vw^{(t-1)}||^2$ is greater than $\epsilon$  \textbf{do} \\
		\normalsize At iteration $t$, \\
		\normalsize 1: $\eta_j^{(t)} \leftarrow  \log \bigg \{a_\gamma + \vone^T \vw_{-j}^{(t-1)} \bigg \} - \log \bigg \{ b_\gamma (r, \kappa) + p - \vone^T\vw_{-j}^{(t-1)} - 1 \bigg \} - \tfrac{1}{2} \log (n+1) + \tfrac{1}{2} (n+1) \log ( \widehat{\sigma}_{j}^2 )$ \\
		\normalsize \hspace{13mm} $- \tfrac{1}{2} (n+1) \log ( \widehat{\sigma}_{j1}^2 )$ \\
		\normalsize 2: $w_j^{(t)} \leftarrow  \text{expit} (\eta_j^{(t)})$ \\
		\normalsize Upon convergence of $\vw$, compute for $i = 1, \ldots, m$ \\
		\normalsize 3: $\widetilde{y}_i \leftarrow \expit \left [ \log \big ( \frac{n_1 + a_y} {n_0 + b_y} \big ) + (1 + \tfrac{1}{n}) \big (\widehat{\vmu}_0 - \widehat{\vmu}_1 \big )^T \mW {\mSigma}^{-1} \big \{\vx_{n+i} - \tfrac{1}{2} (\widehat{\vmu}_0 + \widehat{\vmu}_1)  \big \} \right ] $\\[0.15in]
		\hline
	\end{tabular}
	\\[0.1pt]
\end{table}
\egroup

\noindent where $\log \big \{ (n+1) \widehat{\vsigma}_1^2 \big \}$ is the
element-wise log of the column vector $(n+1) \widehat{\vsigma}_1^2$. Note that
each element in $\widehat{\vsigma}_1^2$ is function of $y_{n+1}$.

Thus, the corresponding variational parameter $\widetilde{y}$ is 
\begin{align*}
\widetilde{y} &= \frac{q(y_{n+1} = 1)}{q(y_{n+1} = 1) + q(y_{n+1} = 0)}, \\
&\approx \expit \left [ \log \bigg ( \frac{n_1 + a_y}{n_0 + b_y} \bigg ) + (1 + \tfrac{1}{n}) \text{ LDA} (\vx_{n+1}) \right ],
\end{align*}
where
\begin{align*}
&\text{ LDA} (\vx_{n+1}) \\ 
&\hspace{8mm} =  \big (\widehat{\vmu}_0 - \widehat{\vmu}_1 \big )^T \mW {\mSigma}^{-1} \big \{\vx_{n+1} 
- \tfrac{1}{2} (\widehat{\vmu}_0 + \widehat{\vmu}_1)  \big \}
\end{align*}
is the na\"{i}ve Bayes LDA classification rule assuming a balanced training dataset that is 
downweighted by $\mW = \text{diag} (w_1, \ldots, w_p)$, and
$\mSigma$ is a diagonal matrix with entries $\{ \widehat{\sigma}_{j1}^2 \}_{j=1}^p$. Note that when
$\mW = \mI_p$ and $a_y=b_y=0$ the classification of
$y_{n+1}$ is the same as for a frequentist na{\"i}ve Bayes LDA.

In the general case
whereby there are $m$ new observations to be classified, the variational parameter for $y_{n+i}$ is
\begin{align}
\label{Yupdate}
\widetilde{y}_i &= \expit \Bigg [ \log \bigg \{ \frac{n_1 
	+ \bE_{-q_i} (\vy_\text{\tiny n+1:n+m}) + a_y} {n_0 + m -  \bE_{-q_i} (\vy_\text{\tiny n+1:n+m}) + b_y} \bigg \} \nonumber \\ 
& \hspace{15mm}+ (1 + \tfrac{1}{n}) \text{ LDA} (\vx_{n+i}) \Bigg ],
\end{align}
where $\bE_{-q_i}$ denotes the expectation with respect to approximate densities of $\vy_{-(n+i)}$ and $\vgamma$. For computational efficiency, 
we may replace the moment estimates in the expression for $\text{ LDA} (\vx_{n+1})$ with their respective\\

\newpage
\hspace{0.1cm}
\vskip 8.25cm

\noindent Taylor's expansion approximation in (\ref{TaylorExp}) and omit the term $\bE_{-q_i} (\vy_\text{\tiny n+1:n+m})$. 
Hence, for $1 \le i \le m$, the updates in (\ref{Yupdate}) may be approximated as
\begin{align}
\label{YupdateTaylors}
\widetilde{y}_i &\approx \expit \Bigg [ \log \bigg ( \frac{n_1 + a_y} {n_0 + b_y} \bigg ) + (1 + \tfrac{1}{n}) \text{ LDA} (\vx_{n+i}) \Bigg ].
\end{align}

\noindent This approximation allows us to update $w_j$ in each RCVB cycle and use only their converged values to
update $\widetilde{y}_{i}$ for $1 \le i \le m$. In addition, the updates of $\widetilde{y}_{i}$ do not depend on 
each other. Hence, the length of each update cycle is kept at $p$. We may use $\widetilde{y}_i$ to construct 
a classification rule by classifying observation $n+i$
to group 1 if $\widetilde{y}_{n+i} > c_y$ for some $c_y \in (0,1)$. 

The RCVB algorithm may be found Table \ref{Algo1}.
Notice that at iteration $t$, we use values of $\vw$ at iteration $t-1$ instead of the values at the current time step. 
This is in line with the batch update strategy described in \cite{Zhang2017}.

	It is worth noting that the appearance of the terms  $\lambda_{\text{LRT}} (\widetilde{\vx}_j)$ and $\text{ LDA} (\vx_{n+i})$ in the updates for 
$w_j$ and $\widetilde{y}_i$ provides reassurance that there is some concordance between the frequentist
methods and our proposed variable selection and classification rules, and hence fulfils the
consistency criterion which we considered in Section \ref{Priors} when
choosing our priors.

\subsection{Keeping the computational cost low}
As mentioned in Section \ref{intro}, a major objective
is to propose a classifier that is scalable and has high computational speed
even under limited computing power. We summarise key steps taken to
help achieve our objective below:
\begin{itemize}
	\item Na{\"i}ve Bayes assumption for group-conditional covariances to avoid inverting huge matrices.
	\item Fast approximate posterior inference using a proposed version of variational inference (RCVB) instead of MCMC which is usually slower to converge.
	\item Reduce the number of steps in each RCVB cycle by Taylor's expansion to 
	isolate updates for classification probabilities ($\widetilde{y}_i$) from the updates of 
	the variable selection probabilities ($w_j$).
	\item Fast implementation with \texttt{C++} that is a compiler-based language in contrast to an interpreter-bsaed language
	such as \texttt{R}. Compilers execute repeat loops more efficiently than interpreters. 
	The reader may refer to \cite{Eddelbuettel2013} for a detailed exposition on the computational advantages of compiler-based languages.
\end{itemize}

\section{Variable selection asymptotics}
\label{Asymptotics}

Here, we establish the consistency of the \texttt{VLDA} variable selection rule. Although \cite{WangBlei2018} has demonstrated
the consistency of variational Bayes estimates in their recent work, their result cannot be directly applied
to our setting as they have assumed a fixed model dimension while we have allowed $p_n$ to
grow with $n$.
The statement for consistency of \texttt{VLDA} is equivalent to showing
a desirable asymptotic property, known as \emph{Chernoff-consistency} for
the sequence of tests that favours the hypothesis $\gamma_j = 1$ if $w_j > c_w$. A sequence of
hypothesis tests $\{ T_{j} \}_{j=1}^p$ is Chernoff-consistent if the sum of type I and II probabilities converge to $0$, i.e.,
\begin{equation*}
\sum_{j \in \sJ_0} \bP (\mbox{Reject } H_{0j} \mbox{ with test } T_{j} ) \rightarrow 0,
\end{equation*}
and the sum of type II errors probabilities
\begin{equation*}
\sum_{j \in \sJ_1} \bP (\mbox{Do not reject } H_{0j} \mbox{ with test } T_{j} ) \rightarrow 0,
\end{equation*}
where $\sJ_0 = \{j : H_{0j} \text{ is true} \}$, $\sJ_1 = \{j : H_{1j} \text{ is true} \}$.
For our case, we need to demonstrate the convergence
$$
E^{(t)} = \sum_{j=1}^p \big | w_j^{(t)}(\mX_j) - \gamma_j^* \big | = o_p(1).
$$
as $n$ approaches infinity, where $\gamma_j^*$ is the true value
of $\gamma_j$. The above sum can be broken down into components: the sum of 
variable selection probabilities among noise variables after cycle $t$
\begin{equation*}
e_{0}^{(t)} = \sum_{\ell \in \sJ_{0n} }  w_\ell^{(t)},
\end{equation*}
\noindent and the complement sum of variable selection probabilities among true signals after cycle $t$
\begin{equation*}
e_{1}^{(t)} = \sum_{\ell \in \sJ_{1n} } \{ 1 - w_\ell^{(t)} \},
\end{equation*}
where the subscripts of $\sJ_{1n}$ and $\sJ_{0n}$ emphasizes their dependence on $n$. 
Note that both $p_{0n} = |\sJ_{0n}| \uparrow \infty$ and  $p_{1n} = |\sJ_{1n}| \uparrow \infty$ as $n$ diverges.
We have also used notations superscripted 
with $*$ to denote the true value of the Gaussian parameters in the frequentist sense. For example, $\mu_{jk}^{*}$ 
denotes the actual population mean of variable $j$ for observations from group $k$.

We shall now state Theorem \ref{mainTheorem} in the main paper. Its proof has been provided in Section 2 
of the Electronic Supplementary Material 1.

\begin{theorem}
	\label{mainTheorem}
	Consider $p_n$ vectors of iid random variables $\widetilde{\mX}_j = (X_{1j},\ldots,X_{nj})^T$, where 
	$(X_{ij}, X_{ik})$ are independent for $j\neq k$, and $p_n$ corresponding pairs of distributional hypotheses
	\begin{equation*}
	H_{0j} \colon X_{ij} \stackrel{\mbox{\scriptsize iid}}{\sim} \sN(\mu_j, \sigma_{j}^{2}),\\
	\end{equation*}
	\noindent versus
	\begin{equation*}
	H_{1j} \colon X_{ij} \stackrel{\mbox{\scriptsize iid}}{\sim} \left\{ \begin{array}{ll}
	\sN(\mu_{j1}, \sigma_{j1}^{2}), & \mbox{ if $y_i = 1$; \; and} \\ [1ex]
	\sN(\mu_{j0}, \sigma_{j1}^{2}), & \mbox{ if $y_i = 0$},
	\end{array} \right.
	\end{equation*}
	\noindent  where $y_i, \; 1 \le i \le n$ are binary constants, $\{ \mu_j, \sigma_j^2 \}$
	is the set of model parameters under $H_{0j}$, and $\{ \mu_{j1}, \mu_{j0}, \sigma_{j1}^2 \}$ is
	the set of model parameters under $H_{1j}$. Assume that the true model is
	\begin{equation*}
	X_{ij} \stackrel{\mbox{\scriptsize iid}}{\sim} \left\{ \begin{array}{ll}
	\sN(\mu_{j1}^*, \sigma_{j1}^{* 2}), & \mbox{ if $y_i = 1$; \; and} \\ [1ex]
	\sN(\mu_{j0}^*, \sigma_{j1}^{*2}), & \mbox{ if $y_i = 0$},
	\end{array} \right.
	\end{equation*}
	where $\mu_{j 1}^*$, $\mu_{j 0}^*$, $\sigma_{j 1}^{*2}$ are the constant parameters
	of the true data-generating Gaussian model. Furthermore, assume that the following set of conditions hold:	
	\begin{enumerate}
		\item the maximum of true Gaussian variances $\max_{j \in \sJ_{1n}}  \sigma_{j1}^{*2}$ are finite for all $n \ge 1$,
		\item the total number of variables is bounded as $p_{n} = O \left ( \exp \{ n / \log(n)  \} \right )$,
		\item the minimal difference in true Gaussian means is bounded, i.e. 
		$\min_{j \in \sJ_{1n}} \delta_j^{*2} = \Omega ( \{\log (n) \}^{-\beta}) $ for some $\beta < r$, where $\delta_j^* =  \lvert \mu_{j1}^* - \mu_{j0}^* \rvert$.
	\end{enumerate}
	
	\noindent The sequence of test that favours $H_{1j}$ (identifies variable $j$ as signal) if $w_j^{(\tau)} > c_w$ 
	for any $c_w \in (0,1)$ and number of cycles $\tau \geq 1$ is Chernoff-consistent, where $\{w_j^{(\tau)} \}_{j=1}^{p_n}$ are obtained
	from the algorithm in Table 1.
\end{theorem}

The above theorem demonstrates the ability of the variable selection rule
to avoid type I error inflation due to the increase of $p_n$, a tendency
to select all true signal variables, and omit all noise variables. However, these properties are only
guaranteed when we choose the cake priors for the Gaussian parameters 
and the prior for $\rho_{\gamma}$ in accordance with Section \ref{Priors}.
We also make a disclaimer that our algorithm may not be optimal with respect to type I and II errors
in any sense.

Implicitly speaking, our resultant variable selection rule is justified
by the asymptotic error rates that they induce. This differs from the existing variable selection 
rules such as higher criticism \citep{DonohoJin2008}
and false-discovery rate controlling  procedures \citep{Benjamini1995}, 
whereby they are justified on the basis of a minimised type II error rate for a 
user-specified family-wise type I error rate and finite sample size $n$.

Our asymptotic justification is in line with theoretical analysis
by \cite{Mai2012} that demonstrated the ability of their model to select
the true discriminative set. Important differences in the assumptions made are summarised in Section 4 of
Electronic Supplementary Material 1. A key advantage
in \cite{Mai2012} is that they have proved their convergence in a setting where
the true group-conditional covariance matrices are not necessarily diagonal, whereas independence between
variables is required in our proof. However, a necessary sparsity condition has been imposed on their discriminant vector \cite{Mai2012}, 
whereas convergence holds in our proof without requiring any sparsity, i.e. $0 < p_{1n} / p_n < 1$ for all $n$.


\section{Numerical results}
\label{Numericals}
In this section, our primary objective is to assess the performance of VaDA classifiers with sixteen simulation settings and six publicly 
available gene expression datasets. A follow-up investigation into the robustness of \texttt{VLDA} to the violation in
variance homogeneity will also be described later. Both \texttt{VLDA} and \texttt{VQDA} variants will be compared 
with other discriminant analysis models. Four of the models are implemented in the R package \url{HiDimDA} \citep{HiDimDA}: 
the diagonal linear discriminant analysis with expanded higher criticism (EHC) - \texttt{Dlda},
the factor model linear discriminant analysis with EHC - \texttt{RFlda}, the maximum uncertainty linear discriminant analysis - \texttt{Mlda}, and 
the shrunken linear discriminant analysis - \texttt{Slda}. 

Other DA models include the nearest shrunken centroid (\texttt{NSC}) classifier
from \url{pamr} \citep{NSC}, lasso (\texttt{penLDA-L1}) and fused-lasso (\texttt{penLDA-FL}) penalised Fisher's linear discriminant analysis from the R package \texttt{penalizedLDA} \citep{penalizedLDA}, the factor-adjusted shrinkage discriminant analysis 
from the package \texttt{FADA} \citep{FADA}, the shrunken centroid linear discriminant analysis from the package \texttt{rda} \citep{Guo2018}, and the sparse linear discriminant analysis in
\texttt{sparseLDA} \citep{Clemmensen2013b}. 

Lastly, we benchmarked these DA models with two presently popular classifiers - the
support vector machine \texttt{svm} \citep{Cortes1995} 
from the package \texttt{LiblineaR} \citep{Helleputte2017} and random forest \citep{Breiman2001} from the package \texttt{caret} \citep{Kuhn2019}. 
The following classifiers \texttt{Dlda-EHC}, \texttt{NSC}, \texttt{penLDA-L1}, \texttt{penLDA-FL}, and our proposed VaDA models
assumed pairwise independence between variables.

The codes for both variants of VaDA have been made publicly available at:
\medskip

\url{http://www.maths.usyd.edu.au/u/jormerod/}

\medskip
\noindent We have also found that a setting of $r = 0.98$ and $\kappa = 10^{-3}$ works well in 
both the simulated and gene expression datasets. The variable selection and classification
thresholds are $c_w = 0.5$ and $c_y =0.5$.
\subsection{Competing classifiers}
\label{competing}
\texttt{Dlda}, \texttt{RFlda}, \texttt{Mlda}, \texttt{Slda} and \texttt{FADA} are 
two-stage classification algorithm that perform variable selection 
and classification in un-integrated stages. In the first stage 
of \texttt{FADA}, the matrix of variables $\mX$ is de-correlated with 
a method described in \cite{Friguet2009}, and is followed by an application 
of the higher criticism threshold \citep[see][]{DonohoJin2008} to select the 
discriminative variables. A modified version of the higher criticism, adapted for large or 
moderate signals, is used in \texttt{Dlda}, \texttt{RFlda}, \texttt{Mlda}, and \texttt{Slda} \citep[see][]{Pedro}.

In the second stage, \texttt{Dlda} fits the na\"{i}ve Bayes LDA model 
with the subset of selected variables whereas \texttt{RFlda} 
fits a factor-based linear discriminant analysis with user-specified $q$ as the 
number of factors. In our analysis, we have chosen the value of $q$ from 
$\lbrace 1, 2, 3\rbrace$ that minimises CV error. The maximum uncertainty LDA \citep{Thomaz2006} that
replaces small eigenvalues of the sample covariance matrix with the mean eigenvalue is fitted in \texttt{Mlda}
and a data-dependent penalised covariance matrix LDA is used in \texttt{Slda}. 
In \texttt{FADA}, the selected variables
from stage one are used to fit a shrinkage 
discriminant analysis model \citep{Ahdesmaki}. Both \texttt{Mlda} and \texttt{Slda}
are good algorithms for benchmarking as they performed excellently on several genomics datasets \citep{Xu2009}. 

\texttt{NSC} classifies observations according to the nearest (group) shrunken centroid. 
A tuning parameter $\Delta$ controls the amount of shrinkage, and consequently the sparsity 
of the estimated difference between the two group centroids. This filters individually weak but 
collectively strong false signals. In our analysis, the adaptive choice of $\Delta$ is adopted 
following procedures in \cite{TibshiraniNSC}. \texttt{NSC} is widely used in biomarker discovery
 \citep[see][]{Liu2005,CraigShapiro2011} and is therefore a good competing classifier.

Lastly, there has been a huge amount of attention given to sparse discriminant analysis models 
in the literature \citep[see][]{Clemmensen2011}. Hence, we have also chosen to compare our 
classifiers with the two versions of \texttt{penLDA} that extended the solution to Fisher's discriminant 
problem \citep{Fisher1936} to high dimension settings. In the two-groups version of
Fisher's discriminant problem, one obtains the estimated discriminant 
vector $\widehat{\vbeta}$ such that
\begin{equation} \label{FisherDA}
\widehat{\vbeta} = \argmax_{\vbeta \in \mathbb{R}^p} (\vbeta^T \widehat{\mSigma}_b \vbeta),
\end{equation}
subject to $\vbeta^T \widehat{\mSigma}_w \vbeta = 1$, where we 
have assumed that the estimated within-class covariance matrix $\widehat{\mSigma}_w$
is full rank and $\widehat{\mSigma}_b$ is the estimated between-class covariance 
matrix. A new observation, $\vx_{n+1}$, is mapped under the transformation 
$\vx_{n+1}^{T} \widehat{\vbeta}$ 
and classified in accordance to its nearest transformed centroid. 
In the \texttt{penLDA} extension, $\widehat{\mSigma}_w$ is replaced with the 
diagonal estimate. Sparsity is induced on $\widehat{\vbeta}$ by imposing either the lasso penalty (\texttt{penLDA-L1}) 
or the fused-lasso penalty (\texttt{penLDA-FL}) on the objective function in (\ref{FisherDA}).
The fused lasso penalty assumes linear ordering in variable indices and is therefore not 
suitable for the publicly available datsets where such ordering cannot be ascertained. 
Tuning parameters are selected from $10^{-4}$, $10^{-3}$, $10^{-2}$, $0.1$ , $1$ or 
$10$ to minimise CV error.

The DA models \texttt{rda} \citep{Guo2007} and \texttt{sparseLDA} \citep{Clemmensen2011}
are examples of non-na{\"i}ve Bayes sparse DA. 
In \texttt{sparseLDA}, the classification problem is re-formulated as a linear least squares problem. Two additional terms are introduced into the resultant objective function to impose
sparsity of both the discriminant vector and the precision matrix estimate. 
The $L_1$ tuning parameter is selected from $10^{-8}$, $10^{-6}$, $10^{-4}$, $10^{-3}$, $0.1$ or $1$ to minimise CV error.
In \texttt{rda}, the group-specific centroids
are replaced with shrunken centroids \citep{TibshiraniNSC} and a penalised covariance matrix estimate with user-specified penalty
is used to replace the MLE. Tuning parameters $\alpha$ and $\delta$ are selected from the intervals $[0,1)$ and $[0,3]$ respectively to minimise CV error.

\subsection{Simulation setting}
\label{sec:Simulation}
We assess the performance of both \texttt{VLDA} and \texttt{VQDA} with
\noindent simulated data. The objective of this simulation study is to identify 
situations in which VaDA's performance differ from the competing classifier. More specifically,
we want to identify the sparsity conditions on the mean differences and the correlation threshold
for our proposed na{\"i}ve Bayes models to work well.
Performance of all classifiers will be assessed with sixteen simulation settings. 
The settings are motivated from those examined in \cite{WittenTibs} and \cite{Clemmensen2013}. The response 
values are generated as follow. We set $y_i$, for each $i$, to take values $1$ or $0$ 
with probability $0.5$. This simulates a balanced study design and also gives adequate 
chance of having at least two observations per group in each simulation repetition. 
The variables are generated as follows.

\begin{align*}
&\vx_{i} | y_i \sim \left\{
\begin{array}{ll}
\sN_p(\vmu_{1}^*,\mSigma^*) , & \text{if } y_i = 1; \; \text{and}, \\ [2ex]
\sN_p(\vmu_{0}^*,\mSigma^*), & \text{if } y_i = 0.
\end{array}
\right. \\
\end{align*}

We use the term \emph{signal strength}, denoted by $d$, to describe the absolute 
standardised difference in the group-conditional means, i.e., 
$d_j = 2|\mu_{j1}^* - \mu_{j0}^*|/(\sigma_{j1}^{*2} + \sigma_{j0}^{*2})$. Each simulation
yields a dataset of $p=500$ variables. 

All simulation settings are homogenous with respect to the group-conditional
variance. For $s = 1,2,3,4$, the simulation settings $\{ s, s+4, s+8, s+12 \}$ have identical specifications for $\vmu_{1}^*$ and $\vmu_{0}^*$
but differ in the specification for $\mSigma^*$. In simulation 1, we specify an independence structure
$\mSigma^* = \text{diag} (\sigma_{1 1}^{*2}, \ldots, \sigma_{p1}^{*2})$. In simulation 5, we simulate a weak correlation structure
where $\mSigma^*$ is specified such that we have 5 networks of $100$ variables each having an $AR(1)$ correlation
where $\rho=0.6$. In simulation 9, we have a moderate correlation structure
where $\mSigma^*$ is specified such that all variables have a $AR(1)$ correlation
where $\rho=0.9$. In simulation 13, we specify a strong correlation structure
where $\mSigma^*$ equals to a uniform correlation matrix with $\rho = 0.8$.

Details of their signal strength settings for simulations 1 to 4 are as follow.\\

\noindent Simulation 1: \\
Here, we simulate the situation when there is a moderate proportion of variables (10\%) 
with non-zero true signals and that the signal strengths ($d_j = 0.7$) are moderately strong. 
We set $\mu_{jk}^* = 0.7$ if $k=1$ and $1 \le j \le 50$ and $\mu_{jk}^* = 0$ otherwise. 
There is also some linear ordering in this setting. This is a homogeneous variance setting 
such that $\sigma_{j1}^{*2} = 1$ for all $1 \leq j \leq p$. \\

\noindent Simulation 2: \\
We test the proposed classifiers in the situation when there is a sizeable proportion (20\%) 
of variables with non-zero true signals and that the signal strengths ($d_j = 0.3$) are weak. We set 
$\mu_{jk}^* = 0.3$ if $k=1$ and $1 \leq j \leq 100$. There is also some linear ordering in 
this setting. This is a homogeneous variance setting such that 
$\sigma_{j1}^{*2} = 1$ for all $1 \leq j \leq p$. \\

\noindent Simulation 3: \\
We test the proposed classifiers in the situation when there is a large proportion (40\%) 
of variables with non-zero true signals and that the signal strengths ($d_j = 0.7$) are moderate. We set 
$\mu_{jk}^* = 0.7$ if $k=1$ and $1 \leq j \leq 200$. There is also some linear ordering in 
this setting. This is a homogeneous variance setting such that 
$\sigma_{j1}^{*2} = 1$ for all $1 \leq j \leq p$. In this simulation setting, we expect all classifiers
to perform well due to the abundance of detectable signals. \\

\noindent Simulation 4: \\
We simulate the situation when we have a very small proportion (2\%) of truly discriminative 
variables and that the non-zero signal strengths each have moderate chance ($\sim$37.0\%) of being moderately strong ($|d_j| \approx 0.6$). 
We let $\mu_{j1}^* \sim \sN(0.5,0.3^2)$ and $\mu_{j0}^* = 0$ for $1 \leq j \leq 10$. 
Otherwise, $\mu_{jk}^* = 0$.  This is a homogeneous variance setting such that
$\sigma_{j1}^{*2} = 1$ for all $1 \leq j \leq p$. \\

\subsection{Performance metrics for simulated datasets}
\label{sec:SimulationMetrics}
The distribution of classification errors for each simulation setting is summarised over 25 simulation 
repetitions with $n = 100$. In each repetition, we generated the $1200$ observations. 
The first $100$ observations $\lbrace (\mathbf{x}_i, y_i) \rbrace_{i=1}^{100}$ are assigned as the 
training set. The next 100 observations $\lbrace (\mathbf{x}_i, y_i) \rbrace_{i=101}^{200}$ are 
designated as the validation set for choosing optimal tuning parameters in $\texttt{penLDA-L1}$ 
and $\texttt{penLDA-FL}$. The remaining $1000$ observations 
$\lbrace (\mathbf{x}_{i}   , y_{i}) \rbrace_{i=201}^{1200}$ will make up the testing dataset. 
At each iteration, the classification error is computed as
\begin{equation}
\text{Classification Error} = \frac{\sum_{i=201}^{1200} I \lbrace y_i \neq \widehat{y}_i \rbrace}{m},
\end{equation}
where $\widehat{y}_i$ is the predicted classification of testing data $i$ and 
$m = 1000$.

Variable selection performance are compared using Matthew's correlation coefficient 
\citep{Matthews}, computed as
\begin{align}
&\text{MCC} \nonumber \\ 
&= \frac{\text{TP} \times \text{TN} - \text{FP} \times \text{FN}}
{\sqrt{(\text{TP} + \text{FP})(\text{TP} + \text{FN})(\text{TN} + \text{FP})(\text{TN} + \text{FN})}},
\end{align}
where TP, TN, FP and FN is the number of true positives, true negatives, false positives 
and false negatives respectively in a particular repetition. A higher value of MCC indicates
better variable selection performance. The MCC presents as a 
suitable variable selection metric for our simulation settings
as it accounts for the imbalance in the total number of truly discriminative (TP $+$ FP)
and non-discriminative (TN $+$ FN) variables \citep{Chicco2017}. Since computational cost has 
also been raised in Section \ref{intro} as a problem with existing models, we shall 
compare the computation time required by each classifier.

\begin{SCfigure*}
	\caption{Classification errors for simulations 1 to 8 ($n=100$). Na{\"i}ve Bayes methods in {\color{darkgreen} green}.}
	\includegraphics[width=0.8\textwidth]{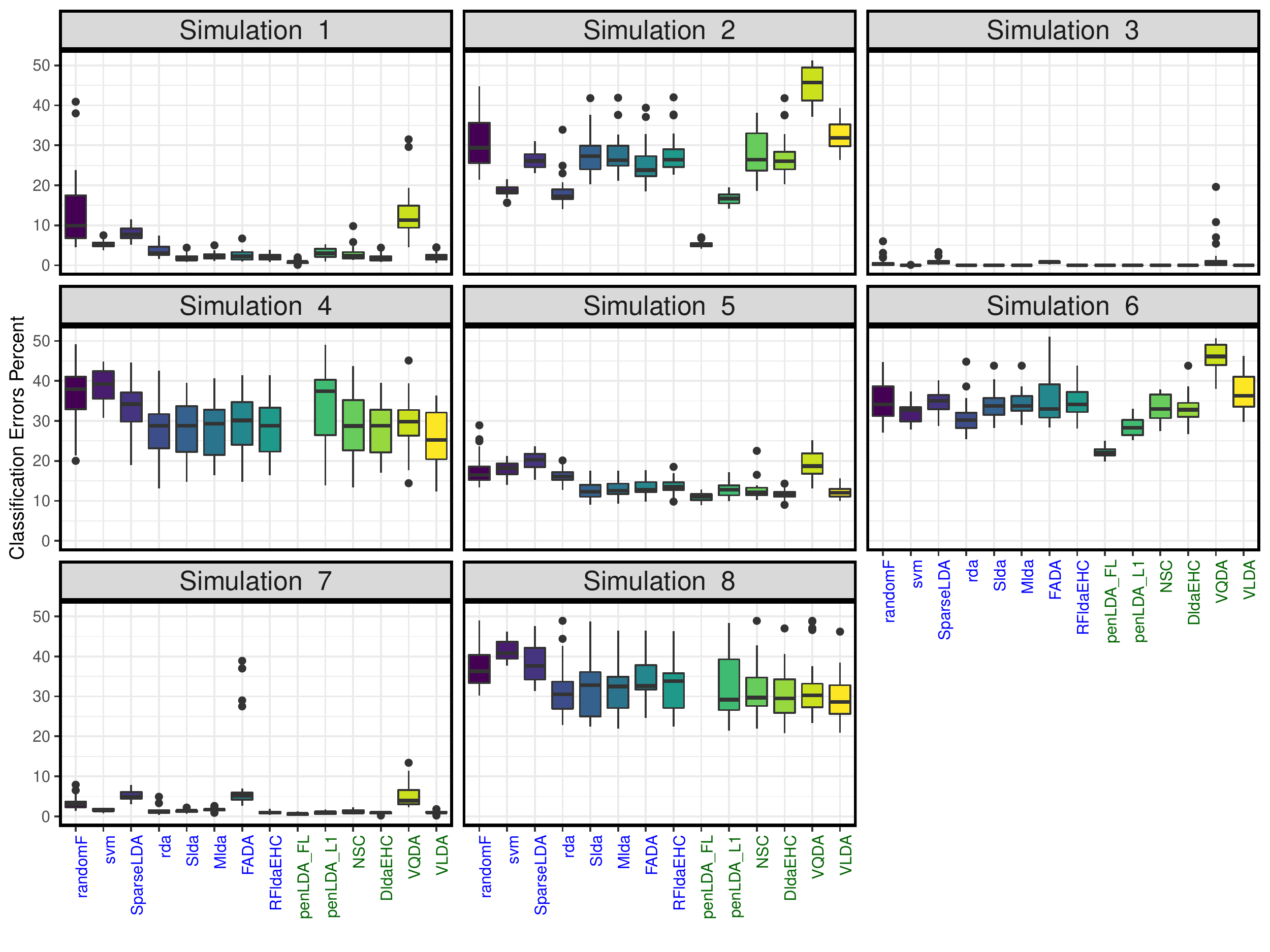}
	\label{fig:1}       
\end{SCfigure*}
\begin{SCfigure*}	
		\includegraphics[width=0.8\textwidth]{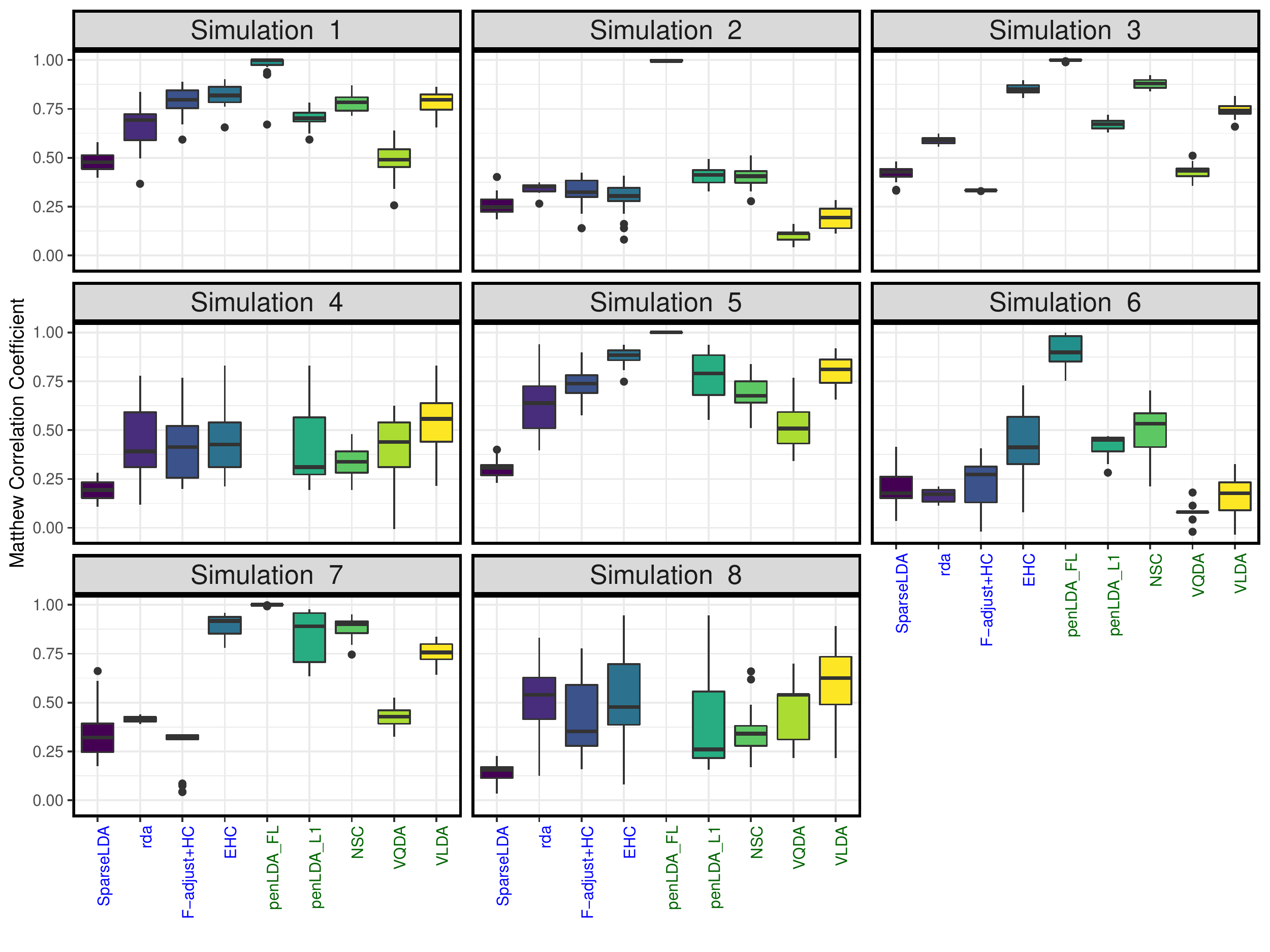}
		\caption{Matthews correlation for simulations 1 to 8 ($n=100$). Na{\"i}ve Bayes methods in {\color{darkgreen} green}.}
	\label{fig:2}       
\end{SCfigure*}

\begin{SCfigure*}
	\includegraphics[width=0.7\textwidth]{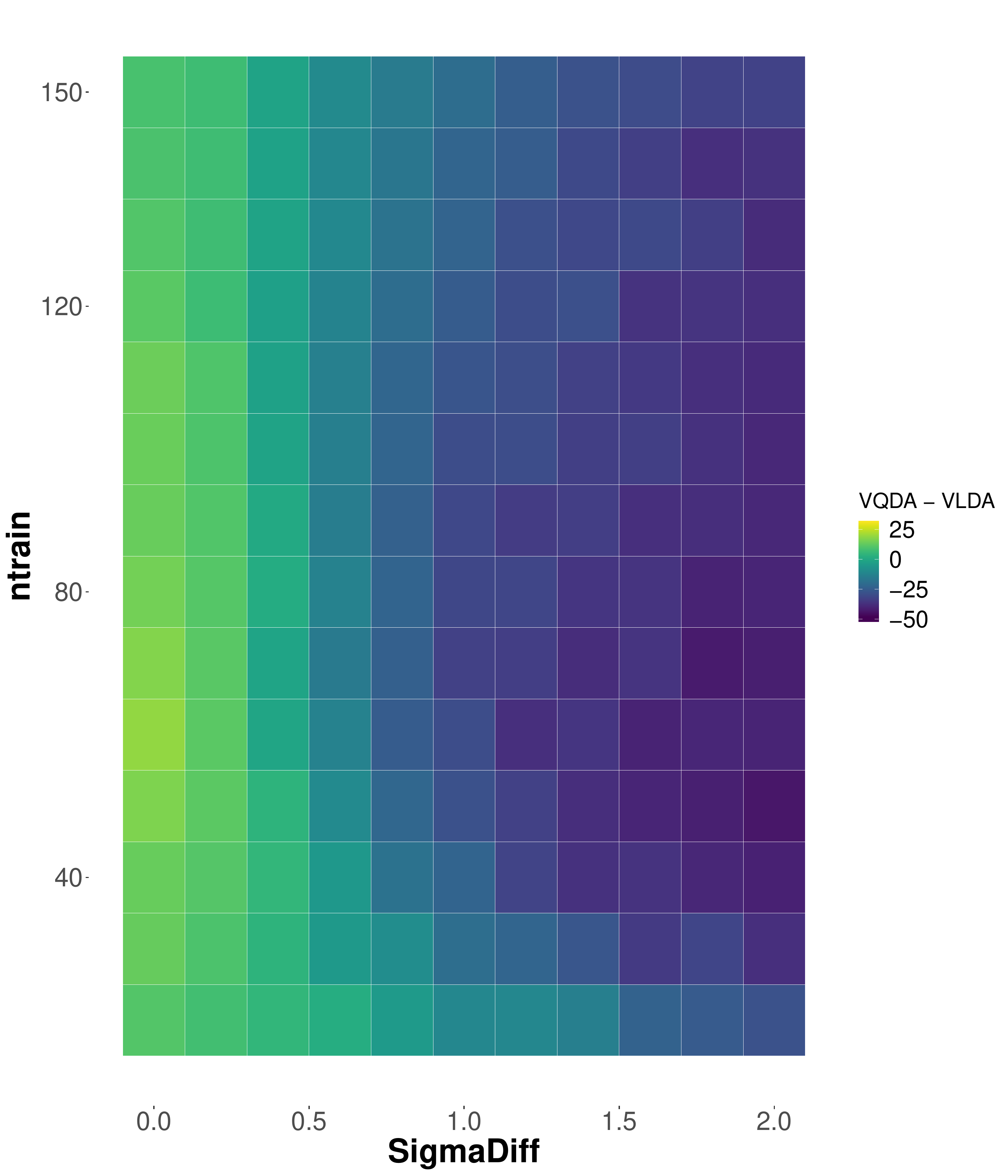}
	\caption{Difference in classification errors (VQDA $-$ VLDA) for varying $n$ and $\Delta_\sigma$ over 25 reps. ($p_1 = 25$)}
	\label{fig:3}       
\end{SCfigure*}

\subsection{Simulation results}
\label{sec:SimResults}
Boxplots of classification errors and MCCs for simulations 1 to 8 may be found in Figures \ref{fig:1} and \ref{fig:2}.
Summary statistics of all simulations are provided in Tables 1 to 4 in Electronic Supplementary Material 2.
Since \texttt{penLDA-FL} requires the variables to have a linear ordering structure, we have left it out of the comparisons for
the simulations setting with randomly drawn means (4, 8, 12 and 16).

\texttt{VLDA} achieved comparatively good classification error performance
under the independence (average rank $=6.3$) and local $AR(1)$ correlation structures (average rank $=5.5$). However, the classifier
performed rather poorly under the global correlation (average rank $=7.3$) and uniform correlation structure (average rank $=11.5$).
Therefore, it is evident that \texttt{VLDA} is robust to a mild violation of the na{\"i}ve Bayes assumption but is
less reliable for classifying data with moderate or strong correlation structure.

Both proposed models also seem to perform relatively better than all other classifiers when the true mean differences
is sparse, i.e. $p_1 / p \le 10 \%$ (simulations 1, 4, 5 and 8) under independent and weak correlation. This is due to the strong global penalty $b_\gamma$ imposed
on all variable selection probabilities. When the signal strengths are weak (simulations 2 and 6),
\texttt{VLDA} did not perform as well in classification error. This may be attributed to the corresponding poor performance
in variable selection. Its poor ability to identify weak signals is explained by observing that the cake priors for the
mean difference $\mu_{j 1} - \mu_{j0}$ is a scale mixture of normals with the log-normal hyperprior on the scale parameter. Although
the log-normal distribution is regarded as heavy-tailed in some textbooks, its tail mass is not heavy
enough to facilitate the weak signals to overcome the $b_\gamma$-global penalty effect. An example of a hyperprior
that preserves weak signals from being masked by the global penalty is the half-Cauchy distribution as reported by \cite{Carvalho2010}.
In contrast, \texttt{penLDA-L1} performed very well in simulations 2 and 6 due to ability of lasso estimators in picking up weak signals as explained in \cite{Fan2010}.

As for variable selection performance in non-weak signal settings, \texttt{VLDA} performed robustly well even under moderate or
strong correlation structure. It is notable that the good performance in variable selection does not translate to good classification performance.
This may be in line with the observations made by \cite{Mai2012} that distinguished between signal variables and discriminative variables. Through
several examples, the authors showed that it is the identification of discriminative variables, and not signal variables, that lead to better classification performance
for LDA models.

%

By changing the group-conditional variances in each setting to create heterogeneous variance settings, 
\texttt{VQDA} outperformed
\texttt{VLDA} and yielded lower classification error than the other classifiers for independent and weak correlation structure (results omitted from paper). This finding
leads us to the follow-up question: how large should the difference in group-conditional variances be
for us to choose \texttt{VQDA} over \texttt{VLDA} as our preferred model? 
We performed further simulations to compare our two proposed models under various regimes of
difference in group-conditionals SDs $\Delta_\sigma$ and training sample size $n$ over 25 repetitions.
Details on the simulation settings and results are summarised in Section 1 of Electronic Supplementary Material 2 and Figure \ref{fig:3}
of this manuscript. Based on Figure \ref{fig:3}
we found that \texttt{VLDA} performs better than \texttt{VQDA} when $\Delta_\sigma \le 0.4$. However, as $n$ increases, the required
$\Delta_\sigma$ for \texttt{VQDA} to perform better than \texttt{VLDA} decreases. When $\Delta_\sigma \ge 1.6$, \texttt{VQDA} performs better than \texttt{VLDA} even for small $n$ (dark blue patches on the right of each panel). A similar pattern in classification error differences is observed
for any $p_1 \ge 10$.
These findings
are similar to those reported in \cite{Marks1974} and \cite{Zavorka2014}. Based on this comparison, 
one should consider the severity of violation in the homogeneity of variance assumption and the training sample size when choosing
between \texttt{VLDA} or \texttt{VQDA}.

\begin{SCfigure*}
	\includegraphics[width=0.8\textwidth]{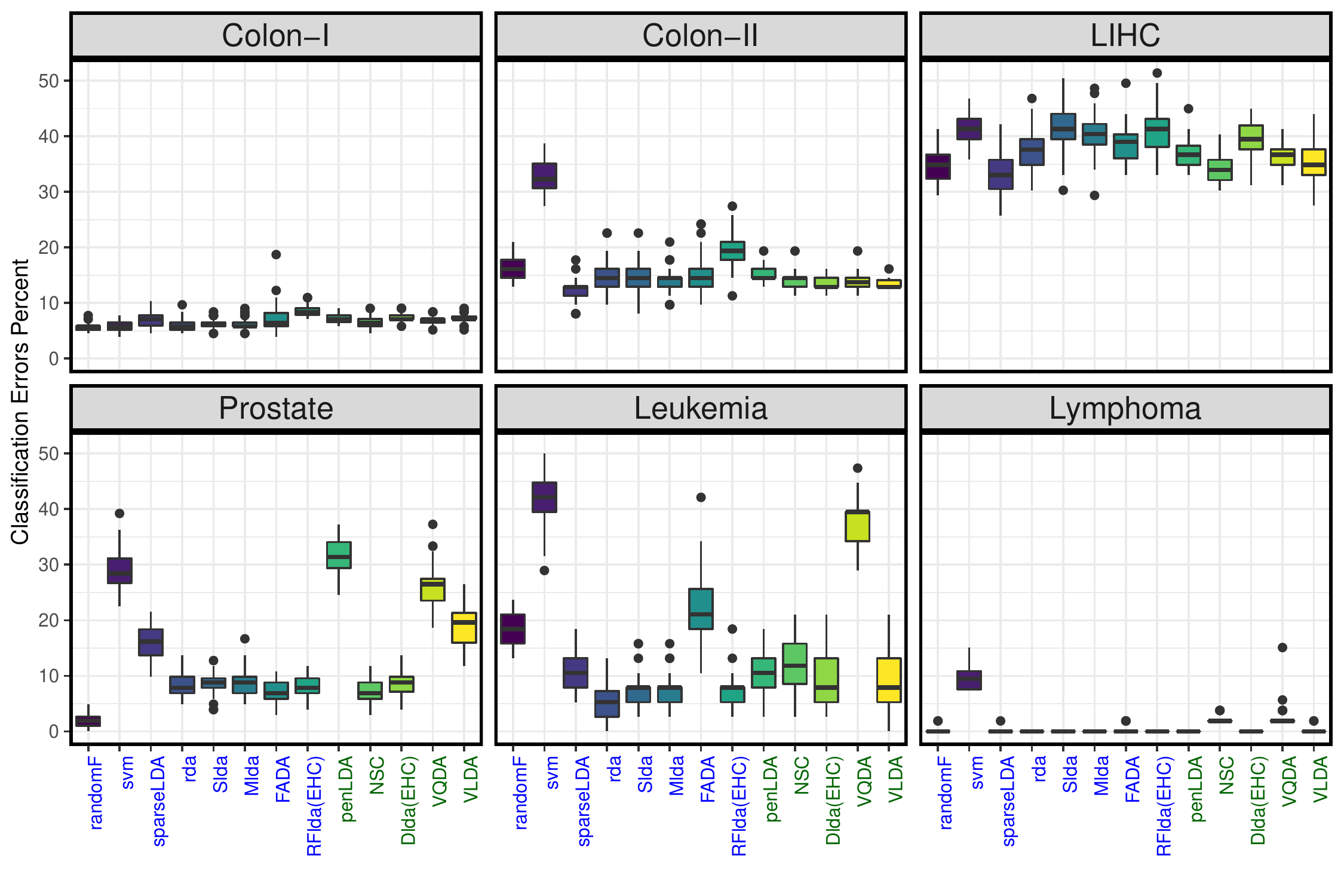}
	\caption{5-fold CV classification errors for genomics datasets (50 reps). Na{\"i}ve Bayes methods in {\color{darkgreen} green}.}
	\label{fig:4}       
\end{SCfigure*}
\subsection{Gene expression datasets}
\label{sec:Realdata}
The classifiers are also compared using six gene expression datasets. The description of the 
datasets may be found in the rest of this subsection.
Since the subset of truly discriminative variables 
are unknown for each dataset, we shall omit the comparison of the variable selection performance. 
A filtering step is applied to leukemia \citep{Golub1999}, {\tt colon I} \citep{Jorissen}, 
and {\tt TCGA-LIHC} \citep{TCGALIHC} 
datasets to remove genes with mostly $0$ readings. We then standardised 
each of the six datasets to obtain $z_{ij} = (x_{ij} - \widetilde{\mu}_j)/s_{j}$ where $x_{ij}$ is the gene $j$ reading for 
observation $i$, $\widetilde{\mu}_{j}$ is the sample mean of gene $j$ and $s_j$ is the sample standard deviation. 
This standardisation procedure is similar to the one in \cite{Dudoit2002}. A 5-fold cross validation over 50 
repetitions is performed.
 
The total number of misclassifications at each iteration is summed across the 
5 CV sub-iterations to compare performance between the classifiers. The classification errors and computational 
time are presented in Figure \ref{fig:4} here and Table 6 of Electronic Supplementary Material 2  respectively. \\

Colorectal cancer dataset I:
This colon cancer dataset is sourced from \url{Bioconductor} and has been analysed by \cite{Jorissen}. 
The dataset consist of $n=155$ 
observations of Affymetrix oligonucleotide arrays. The response variable is 
whether the tumour exhibited microsatellite instability, among which we have 78 microsatellite instable (MSI) 
tumours and 77 microsatellite stable (MSS) tumours. We implemented a filtering step that excludes genes 
with within-class outliers that are either 3 IQR above or below the median. This leaves us with $p=8212$ genes. \\

Colorectal cancer dataset II:
This colon cancer dataset is made available by \cite{Alon1999}. For convenience, we use the version of the dataset
that is found in the \texttt{rda} package. The dataset consist of $n=62$ 
observations of Affymetrix oligonucleotide arrays. The response variable is cell type, among which
we have 40 tumour cells and 22 normal cells. After several pre-filtering steps by \cite{Guo2007}, 
we have $p=2000$ genes that remain for classification.\\

TCGA-LIHC dataset:
The Cancer Genome Atlas Liver Hepatocellular Carcinoma (TCGA-LIHC) dataset is a collection of 
clinical, genetic and pathological data residing in the Genomic Data Commons (GDC) Data Portal 
and is made publicly available \citep{TCGALIHC}. The data underwent pre-processing to remove 
genes with IQR $\leq 0.3$ and the $\text{log}_2 (1 + \text{FPKM})$ transformation is taken. Patients 
whose survival time are lesser the the 20th percentile are considered as poor prognosis ($n_{\text{poor}} = 47$) 
while patients with survival time greater than 80th percentile are labelled as good prognosis ($n_{\text{good}} = 62$). 
Patients who belong to neither survival category were removed from the analysis. This leaves us with $n=109$ 
observations of $15681$ RNA-seq readings.\\

Prostate cancer dataset:
The prostate cancer dataset is made available by \cite{Singh2002}. For convenience, we use the version of the dataset
that is found in the \texttt{sda} package. The dataset consist of $n=102$ 
observations of Affymetrix oligonucleotide arrays. The response variable is cell type, among which
we have 52 tumour cells and 50 normal cells from tissue samples obtained from patients
treated with radical prostatectomy. After several pre-filtering steps by \cite{Ahdesmaki}, 
we have $p=6033$ genes that remain for classification.\\

Leukemia dataset:
The leukemia dataset is made available by \cite{Golub1999}. For convenience, we use the version of the dataset
that is found in the \texttt{plsgenomics} package. The dataset consist of $n=38$ 
observations of Affymetrix oligonucleotide arrays. The response variable is leukemia type, among which
we have 27 ALL type and 50 AML type leukemia samples. After removing genes with low variation $\leq 0.35$, we have
$p=2440$ genes for classification.\\

Lymphoma dataset:
The lymphoma dataset is made available by \cite{Alizadeh2000}. For convenience, we use the version of the dataset
that is found in the \texttt{spls} package. The dataset consist of 62 samples of 3 variants of lymphoma.
Since our proposed model is suitable only for binary classification, we classify a subset of the data consisting of $n=53$ samples
of which 
42 are DLBCL-type and 11 are CLL-type lymphoma. A total of 
$p=4026$ genes are used for classification.\\

\subsection{Gene expression dataset results}
\label{sec:RealdataResults}


Though \texttt{VLDA} is not the best performing classifier for any of the datasets in terms of classification accuracy, 
its classification errors is within a reasonable range from the best performing ones in all datasets except for prostate cancer.
In fact, it is ranked among the top 4 classifiers for both colon II and Leukemia datasets, bearing in mind that it achieved
this level of classification accuracy at a computational speed of 104 to 867 times faster (based on leukemia dataset) than other the other top classifiers 
(see Table 4 of Electronic Supplementary Material 2). We acknowledge that
the performance of our proposed model is less than satisfactory in the prostate cancer dataset.

\texttt{VLDA} outperformed \texttt{VQDA} in all datasets except for colon I which provides
some evidence for the robustness of LDA methods for analysing gene expression datasets.

The performance of the na{\"i}ve Bayes classifiers are generally close to non-na{\"i}ve Bayes classifiers.
This indicative of the lack of a strong correlation structure in these datasets.

%
%
%

\section{Limitations and conclusion}
We have proposed a fast classifier that integrates two common objectives in high dimensional data analysis:
variable selection and classification. The Bayesian 
framework of the classifier lends a two fold-advantage 
to the classifier, provided priors are chosen according to the
recommended settings in this paper. Firstly, it leads us to a variable selection rule that aligns with the 
multiple hypothesis testing paradigm. Although our algorithm may be not asymptotically optimal, we are 
still able to establish consistency of the variable selection rule under
a non-polynomial growth rate for the number of variables. Secondly, the resultant 
variable selection and classification rules are functions of the their respective frequentist
rules, and hence would show high concordance with frequentist results. The classifier is also capable of yielding variable selection and classification results that are comparable to non-na{\"i}ve Bayes DA models under a weak correlation structure. 
Furthermore, this is obtained at
a very small fraction of the computational cost incurred 
by these other DA models.

The speed of our proposed classifiers positions them as a useful exploratory
analysis tool.
In bioinformatics, such high speed algorithms 
may be deployed to mine through a large number of datasets for the purpose of finding new potential markers.

We made the assumption that the variables
are pairwise independent. This assumption is strong and has two implications. First, the classifier may not yield reliable classification results
for moderately or strongly correlated dataset. If there is a need to classify highly correlated data with VaDA, 
one may employ a de-correlation technique that
preserves the original dimension of the variable matrix $\mathbf{X}$ such as the latent factor
method proposed in \cite{Friguet2009}. A study of our classifiers' performance after an implementation of 
de-correlation techniques is beyond the scope of our paper. Secondly, we have established
our asymptotic results under the condition that   the variables are truly independent.
A full investigation into the asymptotic behaviour of \texttt{VLDA} under a non-independent correlation structure will be pursued as future research.

%
%

\newpage
\bgroup
\def\arraystretch{2.4}
\begin{table} \refstepcounter{table}\label{Algo2}
	\centering
	\begin{tabular}{l}
		\normalsize
		\bf{Table 3}~ \normalfont{Iterative scheme for obtaining the parameters in the optimal densities } $q\*(\boldsymbol{\gamma}, y_{n+1}, \ldots, y_{n+m})$ \normalfont{in \texttt{VQDA}}\\
		\hline
		\normalsize Require: For each $j$, initialise $w_j^{(0)}$ with a number in $[0,1]$.\\
		\normalsize \textbf{while} $||\vw^{(t)} - \vw^{(t-1)}||^2$ is greater than $\epsilon$  \textbf{do} \\
		\normalsize At iteration $t$, \\
		\normalsize 1: $\eta_j^{(t)} \leftarrow   \log (\vone^T \vw_{-j}^{(t-1)} + 1) - \log \big \{ p - \vone^T \vw_{-j}^{(t-1)}  - 1 + b_\gamma (r, \kappa) \big \} + \tfrac{1}{2} \log (\tfrac{n_1 n_0}{2})$ \\
		\normalsize $\; \;  \; \;  \; \; \;\;\;\;\;\;\;\;+ \xi(\tfrac{n_1}{2}) + \xi(\tfrac{n_0}{2}) - \xi(\tfrac{n}{2}) - \tfrac{3}{2} \log(n+1) + \tfrac{1}{2} (n+1) \log(\widehat{\sigma}_j^2) - \tfrac{n_1}{2} \log(\widehat{\sigma}_{j1}^2) - \tfrac{n_0}{2} \log(\widehat{\sigma}_{j0}^2)$\\
		\normalsize 2: $w_j^{(t)} \leftarrow  \text{expit} (\eta_j^{(t)})$ \\
		\normalsize Upon convergence of $\vw$, compute for $i = 1, \ldots, m$ \\
		\normalsize 3: $\widetilde{y}_i \leftarrow \text{expit} \bigg [ \log \big( \tfrac{n_1}{n_0} \big ) 
		+ \vone^T \vw \big \{ \log \Gamma(\tfrac{n_1 + 1}{2}) - \log \Gamma(\tfrac{n_1}{2})
		+ \log \Gamma(\tfrac{n_0 + 1}{2}) - \log \Gamma(\tfrac{n_0}{2})  \big \}$ \\
		$\hspace{2.1cm}+ \tfrac{1}{2} \vw^T \Big \{ \log \vphi (\vx_{n+i}; \widehat{\vmu}_1, \widehat{\vsigma}_{1}^2) - \log \vphi (\vx_{n+i}; \widehat{\vmu}_0, \widehat{\vsigma}_{0}^2) \Big \} \bigg ] $\\[0.15in]
		
		\hline
	\end{tabular}
	\\[0.5pt]
\end{table}
\egroup

Despite its limitations, 
we believe that VaDA is a computationally-efficient option for analysing most high dimensional datasets.

\section*{Acknowledgement}
The authors would like to thank Rachel Wang (University of Sydney), the associate editor, and the anonymous reviewers
for their valuable feedback to improve this manuscript.

\section*{Disclosure of potential conflicts of interest}
Conflict of Interest: The authors declare that they have no conflict of interest.

\section*{Appendix}
\label{appendix}
\subsection*{VQDA derivations}
In the \texttt{VQDA} setting ($\sigma_{j 1}^2 \neq \sigma_{j 0}^2$) the posterior distribution
of $\gamma_j$ given $\sD$ and $\vx_{n+1}$ may be expressed as\\

\begin{equation*}
p(\gamma_j \; | \; \sD, \vx_{n+1}) = \frac{ p(\gamma_j, \sD, \vx_{n+1})}{ p(\sD, \vx_{n+1}) }.
\end{equation*}
\noindent By letting $h \rightarrow \infty$, the marginal likelihood of the data in the denominator is of
the same form as equation (\ref{margLik}) 
\newpage
\hspace{0.1cm}
\vskip 9.2cm
\noindent with the exception that
\begin{equation*}
\vtheta_1 = (\vmu_1, \vmu_0, \vmu, \vsigma_1^2, \vsigma_0^2, \vsigma^2, \rho_y, \rho_\gamma),
\end{equation*}
and
\begin{align*}
\lambda_{\text{LRT}} &(\widetilde{\vx}_j, x_{n+1,j}, \vy, y_{n+1}) \\
&= (n+1) \log(\widehat{\sigma}_j^2) - 
(n_1 + y_{n+1}) \log(\widehat{\sigma}_{j1}^2) \\ & \hspace{0.5cm} - (n_0 + 1 - y_{n+1}) \log(\widehat{\sigma}_{j0}^2),
\end{align*}
\noindent where
\begin{align*}
&\widehat{\sigma}_{j 1}^2 = \tfrac{1}{n_1+y_{n+1}} \bigg [ ||\vy^T\{\widetilde{\vx}_{j} 
- \widehat{\mu}_{j1}\vone\}||^2 \\ &\hspace{2.6cm}+ y_{n+1} (x_{n+1,j} - \widehat{\mu}_{j1})^2 
\bigg ],
\end{align*}
\begin{align*}
&\widehat{\sigma}_{j 0}^2 = \tfrac{1}{n_0 +1 - y_{n+1}} \bigg [ ||(\vone - \vy)^T\{\widetilde{\vx}_{j} 
- \widehat{\mu}_{j0}\vone\}||^2 \\ 
& \hspace{2.7cm} + (1 - y_{n+1}) (x_{n+1,j} - \widehat{\mu}_{j0})^2 \bigg ],
\end{align*}
\noindent and the $j^{\mbox{\scriptsize th}}$ entry of $\vlambda_{\mbox{\scriptsize Bayes}}$ is (as $h \rightarrow \infty$)
\begin{align*}
&\lambda_{\mbox{\scriptsize Bayes}} (\widetilde{\vx}_j, x_{n+1,j}, \vy, y_{n+1}) \\
&\rightarrow \lambda_{\mbox{\scriptsize LRT}}(\widetilde{\vx}_j, x_{n+1,j}, \vy, y_{n+1})
+ \log (n_1 + y_{n+1} ) \\
&\hspace{0.4cm}  + \log (n_0 + 1 - y_{n+1} ) -  \log (2) 
- 3\log(n+1) \\
&\hspace{0.4cm} - 2\xi\{(n+1)/2\} + 2\xi\{(n_1 + y_{n+1})/2\} \\
&\hspace{0.4cm} + 2\xi\{ (n_0 + 1 - y_{n+1})/2 \}, \\
&= \lambda_{\mbox{\scriptsize LRT}}(\widetilde{\vx}_j, x_{n+1,j}, \vy, y_{n+1})
- 2 \log(n+1) \\
&\hspace{0.3cm} + O(n_0^{-1} + n_1^{-1}),
\end{align*}
\noindent where $\xi (x) = \log \Gamma(x) + x - x \log(x) - \tfrac{1}{2} \log(2 \pi)$. 
Since the calculation of the marginal likelihood
involves a combinatorial sum over $2^{p+1}$ binary combinations, exact Bayesian
inference is also computationally impractical in the \texttt{VQDA} setting.

Similar to \texttt{VLDA}, we will use RCVB to approximate the posterior 
$p(\vgamma, y_{n+1} | \vx, \vx_{n+1}, \vy)$ by
\begin{align*}
q(y_{n+1}, \vgamma) = q (y_{n+1}) 
\prod_{j=1}^{p} q_j (\gamma_j).
\end{align*}
\noindent This yields the approximate posterior for $\gamma_j$ as
\begin{align*}
&q_j(\gamma_j) \propto \int \exp \big [ \bE_{-q_j} \{ \log p(\sD, \vx_{n+1}, y_{n+1},\vgamma, \vtheta_1) \}  \big ] \text{d} \vtheta_1, \\
&\propto \exp\bigg[ \bE_{-q_j} \Big \{ \log\sB(a_\gamma + \vone^T\vgamma, b_\gamma + p - \vone^T\vgamma) \Big \} \\ 
&\hspace{12mm}+ \tfrac{\gamma _j}{2} \bE_{-q_j} \Big \{ \vlambda_{\mbox{\scriptsize Bayes}} (\widetilde{\vx}_j, x_{n+1,j}, \vy, y_{n+1}) \Big \} \bigg ].
\end{align*}

\noindent For a sufficiently large $n$, we can avoid the need to evaluate the expectation 
$ \bE_{-q_j} \Big \{ \vlambda_{\mbox{\scriptsize Bayes}} (\widetilde{\vx}_j, x_{n+1,j}, \vy, y_{n+1}) \Big \}$
by applying Taylor's expansion to obtain the approximation
\begin{align}
\label{TaylorExpQDA}
&\bE_{-q_j} \log(a_\gamma + \vone^T \vgamma_{-j}) \approx \log(a_\gamma + \vone^T \vw_{-j}), \nonumber \\
&\bE_{-q_j} \log(b_\gamma + p - \vone^T \vgamma_{-j} - 1) \nonumber \\ 
&\hspace{8mm} \approx \log(b_\gamma + p - \vone^T \vw_{-j} - 1), \nonumber \\
&\widehat{\sigma}_{j 1}^2 \approx \tfrac{1}{n_1} ||\vy^T\{\widetilde{\vx}_{j} - \widehat{\mu}_{j1}\vone\}||^2, \nonumber \\
&\widehat{\sigma}_{j 0}^2 \approx \tfrac{1}{n_0} ||(\vone - \vy)^T\{\widetilde{\vx}_{j} - \widehat{\mu}_{j0}\vone\}||^2, \nonumber\\
&\widehat{\sigma}_{j}^2 \approx  \tfrac{1}{n} ||\widetilde{\vx}_{j} - \widehat{\mu}_{j}\vone||^2, \nonumber \\
&\widehat{\mu}_{j 1} \approx \tfrac{1}{n_1} \vy^T \widetilde{\vx}_j , 
\;\; \widehat{\mu}_{j 0} \approx \tfrac{1}{n_0} (\vone - \vy)^T \widetilde{\vx}_j, 
\;\; \widehat{\mu}_{j} \approx \tfrac{1}{n} \vone^T \widetilde{\vx}_j,
\end{align}
\noindent and, similar to \texttt{VLDA}, $\vlambda_{\mbox{\scriptsize Bayes}}$ does 
not depend on the new observation $(\vx_{n+1}, y_{n+1})$. 
By using the approximation in (\ref{TaylorExpQDA}), we have
\begin{align*}
&w_j = \frac{q_j(\gamma_j = 1)}{q_j(\gamma_j = 1) + q_j(\gamma_j = 0)}, \\
&\approx \expit\bigg[ \log (a_\gamma + \vone^T \vw_{-j}) - \log (b_\gamma + p - \vone^T \vw_{-j} - 1) \\
&\hspace{14mm}+ \tfrac{1}{2} \log (\tfrac{n_1 n_0}{2}) + \xi(\tfrac{n_1}{2}) + \xi(\tfrac{n_0}{2}) - \xi(\tfrac{n}{2}) \\
&\hspace{14mm} - \tfrac{3}{2} \log(n+1) + \tfrac{1}{2} \lambda_{\mbox{\scriptsize LRT}} (\widetilde{\vx}_j, y_{n+1})   \bigg ], \\
&= \expit\bigg[ \text{penalty}_{QDA,j} + \tfrac{1}{2} \lambda_{\mbox{\scriptsize LRT}} (\widetilde{\vx}_j, y_{n+1}) \bigg ],
\end{align*}

To obtain the approximate density for $y_{n+1}$, we integrate analytically over
$\vtheta_1$ to obtain
\begin{align*}
&q(y_{n+1}) \propto \int \exp \big [ \bE_{-y} \{ \log p(\sD, \vx_{n+1}, y_{n+1},\vgamma, \vtheta_1) \}  \big ] \text{d} \vtheta_1, \\
&\propto \exp\bigg[ \log\sB(a_y + n_1 + y_{n+1}, b_y + n_0 + 1 - y_{n+1}) \\ 
&\hspace{12mm} + \vone^T \vw \Big \{ \log \Gamma(\tfrac{n_1 + y_{n+1}}{2}) 
+ \log \Gamma(\tfrac{n_0 + 1 - y_{n+1}}{2}) \Big \}  \\
&\hspace{3mm} + \tfrac{1}{2} \vw^T \Big \{  \log \vphi (\vx_{n+1}; \widehat{\vmu}_1, \widehat{\vsigma}_{1}^2) - \log \vphi (\vx_{n+1}; \widehat{\vmu}_0, \widehat{\vsigma}_{0}^2) \Big \} \bigg ],
\end{align*}
where the $j^{\mbox{\scriptsize th}}$ element of the $p \times 1$ vector
$\vphi (\vx_{n+1}; \widehat{\vmu}_k, \widehat{\vsigma}_{k}^2)$ is the Gaussian density
$$
\phi(x_{n+1,j}; \widehat{\mu}_{j k}, \widehat{\sigma}_{j k}^2),
$$
and the $\log$ prefix denotes an element-wise $\log$ of a vector.

In the general case with $m$ new observations, we may apply Taylor's expansion results
from (\ref{TaylorExpQDA}) to compute the approximate classification probability for $y_{n+i}$ as
\begin{align*}
&\widetilde{y}_i = \frac{q(y_{n+i} = 1)}{q(y_{n+i} = 1) + q(y_{n+i} = 0)}, \\
&\hspace{3mm} \approx \expit \bigg[ \log \big( \tfrac{n_1}{n_0} \big ) 
+ \vone^T \vw \big \{ \log \Gamma(\tfrac{n_1 + 1}{2}) - \log \Gamma(\tfrac{n_1}{2}) \\
&\hspace{35mm}  + \log \Gamma(\tfrac{n_0 + 1}{2}) - \log \Gamma(\tfrac{n_0}{2})  \big \} \\
&+ \tfrac{1}{2} \vw^T \Big \{  \log \vphi (\vx_{n+i}; \widehat{\vmu}_1, \widehat{\vsigma}_{1}^2) - \log \vphi (\vx_{n+i}; \widehat{\vmu}_0, \widehat{\vsigma}_{0}^2) \Big \} \bigg ].
\end{align*}
The RCVB alogrithm for \texttt{VQDA} may be found in Table 3.


\bibliographystyle{spcustom}

\begin{thebibliography}{67}
\providecommand{\natexlab}[1]{#1}
\providecommand{\url}[1]{\texttt{#1}}
\providecommand{\urlprefix}{URL }
\expandafter\ifx\csname urlstyle\endcsname\relax
  \providecommand{\doi}[1]{doi:\discretionary{}{}{}#1}\else
  \providecommand{\doi}{doi:\discretionary{}{}{}\begingroup
  \urlstyle{rm}\Url}\fi
\providecommand{\selectlanguage}[1]{\relax}

\bibitem[{Ahdesm{\"a}ki and Strimmer(2010)}]{Ahdesmaki}
Ahdesm{\"a}ki, M., Strimmer, K.: {Feature selection in omics prediction
  problems using CAT score and false discovery rate control}.
\newblock The Annals of Applied Statistics \textbf{4}(1), 503--519 (2010)

\bibitem[{Alizadeh et~al.(2000)Alizadeh, Eisen, Davis, Ma, Lossos, Rosenwald,
  Boldrick, Sabet, Tran, Yu, Powell, Yang, Marti, Moore, Hudson, Lu, Lewis,
  Tibshirani, Sherlock, Chan, Greiner, Weisenburger, Armitage, Warnke, Levy,
  Wilson, Grever, Byrd, Botstein, Brown and Staudt}]{Alizadeh2000}
Alizadeh, A., Eisen, M., Davis, R., Ma, C., Lossos, I., Rosenwald, A.,
  Boldrick, J., Sabet, H., Tran, T., Yu, X., Powell, J., Yang, L., Marti, G.,
  Moore, T., Hudson, J.J., Lu, L., Lewis, D., Tibshirani, R., Sherlock, G.,
  Chan, W., Greiner, T., Weisenburger, D., Armitage, J., Warnke, R., Levy, R.,
  Wilson, W., Grever, M., Byrd, J., Botstein, D., Brown, P., Staudt, L.:
  Distinct types of diffuse large b-cell lymphoma identified by gene expression
  profiling.
\newblock Nature \textbf{403}, 503--511 (2000)

\bibitem[{Alon et~al.(1999)Alon, Barkai, Notterman, Gish, Ybarra, Mack and
  AJ}]{Alon1999}
Alon, U., Barkai, N., Notterman, D., Gish, K., Ybarra, S., Mack, D., AJ, L.:
  Broad patterns of gene expression revealed by clustering analysis of tumor
  and normal colon tissues probed by oligonucleotide arrays.
\newblock Proceedings of the National Academy of Sciences \textbf{96}(12),
  6745--6750 (1999)

\bibitem[{Benjamini and Daniel(2001)}]{Benjamini2001}
Benjamini, Y., Daniel, Y.: {The control of the false discovery rate in multiple
  testing under dependency}.
\newblock The Annals of Statistics \textbf{29}(4), 1165–--1188 (2001)

\bibitem[{Benjamini and Hochberg(1995)}]{Benjamini1995}
Benjamini, Y., Hochberg, Y.: {Controlling the false discovery rate: a practical
  and powerful approach to multiple testing}.
\newblock Journal of the Royal Statistical Society, Series B \textbf{57}(1),
  289–--300 (1995)

\bibitem[{Bickel and Levina(2004)}]{Bickel}
Bickel, P.J., Levina, E.: {Some theory for Fisher's linear discriminant
  function, `na{\"i}ve Bayes' and some alternatives when there are many more
  variables than observations}.
\newblock Bernoulli \textbf{10}(6), 989--1010 (2004)

\bibitem[{Blei and Jordan(2006)}]{bleiDPmixture}
Blei, D.M., Jordan, M.I.: {Variational inference for Dirichlet processes}.
\newblock Bayesian Analysis \textbf{1}(1), 121--144 (2006)

\bibitem[{Blei et~al.(2017)Blei, Kucukelbir and McAuliffe}]{Blei2017}
Blei, D.M., Kucukelbir, A., McAuliffe, J.D.: {Variational inference: A review
  for statisticians}.
\newblock Journal of American Statistical Association \textbf{112}(518),
  859--877 (2017)

\bibitem[{Bonferroni(1936)}]{Bonferroni1936}
Bonferroni, C.E.: Teoria statistica delle classi e calcolo delle probabilità.
\newblock Pubblicazioni del R Istituto Superiore di Scienze Economiche e
  Commerciali di Firenze (1936)

\bibitem[{Breiman(2001)}]{Breiman2001}
Breiman, L.: Random forests.
\newblock Machine Learning \textbf{45}(1), 5--32 (2001)

\bibitem[{Cai and Liu(2011)}]{CaiLiu}
Cai, T., Liu, W.: {A direct estimation approach to sparse linear discriminant
  analysis}.
\newblock Journal of the American Statistical Association \textbf{106}(496),
  1566--1577 (2011)

\bibitem[{Carvalho et~al.(2010)Carvalho, Polson and Scott}]{Carvalho2010}
Carvalho, C.M., Polson, N.G., Scott, J.G.: {The horseshoe estimator for sparse
  signals}.
\newblock Biometrika \textbf{97}(2), 465--280 (2010)

\bibitem[{Chen and Feng(2014)}]{Chen2014}
Chen, Y., Feng, J.: {Efficient method for Moore-Penrose inverse problems
  involving symmetric structrue based on group theory}.
\newblock Journal of Computing in Civil Engineering \textbf{28}(2), 182--190
  (2014)

\bibitem[{Chicco(2017)}]{Chicco2017}
Chicco, D.: Ten quick tips for machine learning in computational biology.
\newblock BioData Mining \textbf{10}(35), 1--17 (2017)

\bibitem[{Clemmensen(2013)}]{Clemmensen2013}
Clemmensen, L.: On discriminant analysis techniques and correlation structures
  in high dimensions.
\newblock Kgs. Lyngby: Technical University of Denmark (DTU). Technical
  Report-2013 (4) (2013)

\bibitem[{Clemmensen and Kuhn(2016)}]{Clemmensen2013b}
Clemmensen, L., Kuhn, M.: sparseLDA: Sparse Discriminant Analysis (2016).
\newblock R package version 0.1-9

\bibitem[{Clemmensen et~al.(2011)Clemmensen, Witten, Hastie and
  Ersboll}]{Clemmensen2011}
Clemmensen, L., Witten, D., Hastie, T., Ersboll, B.: {Sparse discriminant
  analysis}.
\newblock Technometrics \textbf{53}(4), 406--413 (2011)

\bibitem[{Cortes and Vapnik(1995)}]{Cortes1995}
Cortes, C., Vapnik, V.: Support-vector networks.
\newblock Machine Learning \textbf{20}(3), 273--297 (1995)

\bibitem[{Courrieu(2005)}]{Courrieu2005}
Courrieu, P.: {Fast computation of Moore-Penrose inverse matrices}.
\newblock Neural Information Processing - Letters and Reviews \textbf{8}(2),
  25--29 (2005)

\bibitem[{Craig-Shapiro et~al.(2011)Craig-Shapiro, Kuhn, Xiong, Pickering, Liu,
  Misko, Perrin, Bales, Soares, Fagan and David}]{CraigShapiro2011}
Craig-Shapiro, R., Kuhn, M., Xiong, C., Pickering, E.H., Liu, J., Misko, T.P.,
  Perrin, R.J., Bales, K.R., Soares, H., Fagan, A.M., David, M.H.: {Multiplexed
  immunoassay panel identifies novel CSF biomarkers for Alzheimer’s disease
  diagnosis and prognosis}.
\newblock PLoS ONE p. e18850 (2011)

\bibitem[{Donoho and Jin(2004)}]{DonohoJin2004}
Donoho, D., Jin, J.: {Higher criticism for detecting sparse heterogeneous
  mixtures}.
\newblock The Annals of Statistics \textbf{32}(3), 962–--994 (2004)

\bibitem[{Donoho and Jin(2008)}]{DonohoJin2008}
Donoho, D., Jin, J.: {Higher criticism thresholding. optimal feature selection
  when useful features are rare and weak}.
\newblock Proceedings of the National Academy of Sciences \textbf{105}(39),
  14790--14795 (2008)

\bibitem[{Duarte~Silva(2011)}]{Pedro}
Duarte~Silva, P.A.: {Two group classification with high-dimensional correlated
  data: a factor model approach}.
\newblock Computational Statistics and Data Analysis \textbf{55}(11),
  2975--2990 (2011)

\bibitem[{Duarte~Silva(2015)}]{HiDimDA}
Duarte~Silva, P.A.: HiDimDA: High dimensional discriminant analysis (2015).
\newblock R package version 0.2-4

\bibitem[{Dudoit et~al.(2002)Dudoit, Fridyland and Speed}]{Dudoit2002}
Dudoit, S., Fridyland, J., Speed, T.P.: {Comparison of discrimination methods
  for classification of tumours using gene expression data}.
\newblock Journal of American Statistical Association \textbf{97}(457), 77--87
  (2002)

\bibitem[{Eddelbuettel(2013)}]{Eddelbuettel2013}
Eddelbuettel, D.: Seamless R and C++ integration with Rcpp.
\newblock Springer (2013)

\bibitem[{Erickson et~al.(2016)Erickson, Kirk, Lee, Bathe, Kearns, Gerdes,
  Rieger-Christ and Lemmerman}]{TCGALIHC}
Erickson, B.J., Kirk, S., Lee, Y., Bathe, O., Kearns, M., Gerdes, C.,
  Rieger-Christ, K., Lemmerman, J.: {Radiology Data from The Cancer Genome
  Atlas Liver Hepatocellular Carcinoma [TCGA-LIHC] collection.}
\newblock The Cancer Imaging Archive.  (2016)

\bibitem[{Fan and Fan(2008)}]{FanFan}
Fan, J., Fan, Y.: {High-dimensional classification using features annealed
  independence rules}.
\newblock The Annals of Statistics \textbf{36}(6), 2605--2637 (2008)

\bibitem[{Fan and Lv(2010)}]{Fan2010}
Fan, J., Lv, J.: {A selective overview of variable selection in high
  dimensional feature space}.
\newblock Statistica Sinica \textbf{20}(1), 101--148 (2010)

\bibitem[{Fern\'{a}ndez-Delgado et~al.(2014)Fern\'{a}ndez-Delgado, Cernadas and
  Barro}]{Fernandez-Delgado}
Fern\'{a}ndez-Delgado, M., Cernadas, E., Barro, S.: {Do we need hundreds of
  classifiers to solve real world classification problems?}
\newblock Journal of Machine Learning \textbf{15}, 3133--3181 (2014)

\bibitem[{Fisher(1936)}]{Fisher1936}
Fisher, R.A.: The use of multiple measurements in taxonomic problems.
\newblock Annals of Eugenics \textbf{7}(2), 179--188 (1936)

\bibitem[{Fisher and Sun(2011)}]{Fisher2011}
Fisher, T., Sun, X.: Improved stein-type shrinkage estimators for the
  high-dimensional multivariate normal covariance matrix.
\newblock Computational Statistics and Data Analysis \textbf{55}(1), 1909--1918
  (2011)

\bibitem[{Friedman(1989)}]{Friedman1989}
Friedman, J.H.: {Regularized discriminant analysis}.
\newblock Journal of the American Statistical Association \textbf{84}(405),
  165--175 (1989)

\bibitem[{Friguet et~al.(2009)Friguet, Kloareg and Causeur}]{Friguet2009}
Friguet, C., Kloareg, M., Causeur, D.: A factor model approach to multiple
  testing under dependence.
\newblock Journal of the American Statistical Association \textbf{104}(488),
  1406--1415 (2009)

\bibitem[{Golub et~al.(1999)Golub, Slonim, Tamayo, Huard, Gaasenbeek and
  Mesirov}]{Golub1999}
Golub, T., Slonim, D., Tamayo, P., Huard, C., Gaasenbeek, M., Mesirov, J.:
  Molecular classification of cancer: Class discovery and class prediction by
  gene expression monitoring.
\newblock Science \textbf{286}(5439), 531--537 (1999)

\bibitem[{Guo et~al.(2007)Guo, Hastie and Tibshirani}]{Guo2007}
Guo, Y., Hastie, T., Tibshirani, R.: Regularized linear discriminant analysis
  and its application in microarrays.
\newblock Biostatistics \textbf{8}(1), 86--100 (2007)

\bibitem[{Guo et~al.(2018)Guo, Hastie and Tibshirani}]{Guo2018}
Guo, Y., Hastie, T., Tibshirani, R.: rda: Shrunken Centroids Regularized
  Discriminant Analysis (2018).
\newblock R package version 1.0.2-2.1

\bibitem[{Hastie et~al.(2014)Hastie, Tibshirani, Narasimhan and Chu}]{NSC}
Hastie, T., Tibshirani, R., Narasimhan, B., Chu, G.: pamr: Prediction analysis
  for microarrays (2014).
\newblock R package version 1.55

\bibitem[{Helleputte(2017)}]{Helleputte2017}
Helleputte, T.: LiblineaR: Linear Predictive Models Based on the LIBLINEAR
  C/C++ Library (2017).
\newblock R package version 2.10-8

\bibitem[{Jorissen et~al.(2008)Jorissen, Lipton, Gibbs, Chapman, Desai, Jones,
  Yeatman, East, Tomlinson, Verspaget, Aaltonen, Kruhoffer, Orntoft, Andersen
  and Sieber}]{Jorissen}
Jorissen, R.N., Lipton, L., Gibbs, P., Chapman, M., Desai, J., Jones, I.T.,
  Yeatman, T.J., East, P., Tomlinson, I.P., Verspaget, H.W., Aaltonen, L.A.,
  Kruhoffer, M., Orntoft, T.F., Andersen, C.L., Sieber, O.M.: {DNA copy-number
  alterations underlie gene expression differences between microsatellite
  stable and unstable colorectal cancers}.
\newblock Clinical Cancer Research \textbf{14}(24), 8061--8069 (2008)

\bibitem[{Kuhn et~al.(2019)Kuhn, Wing, Weston, Williams, Keefer, Engelhardt,
  Cooper, Mayer, Kenkel, Benesty, Lescarbeau, Ziem, Scrucca, Tang, Candan and
  Hunt}]{Kuhn2019}
Kuhn, M., Wing, J., Weston, S., Williams, A., Keefer, C., Engelhardt, A.,
  Cooper, T., Mayer, Z., Kenkel, B., Benesty, M., Lescarbeau, R., Ziem, A.,
  Scrucca, L., Tang, Y., Candan, C., Hunt, T.: caret: Classification and
  Regression Training (2019).
\newblock R package version 6.0-84

\bibitem[{Liu et~al.(2005)Liu, Cutler, Li, Pan, Peng, Hoey, Chen and
  Ling}]{Liu2005}
Liu, J.J., Cutler, G., Li, W., Pan, Z., Peng, S., Hoey, T., Chen, L., Ling,
  X.B.: {Multiclass cancer classification and biomarker discovery using
  GA-based algorithms}.
\newblock Bioinformatics \textbf{21}(11), 2691–--2697 (2005)

\bibitem[{Luts and Ormerod(2014)}]{Luts2014}
Luts, J., Ormerod, J.T.: {Mean field variational Bayesian inference for support
  vector machine classification}.
\newblock Computational Statistics and Data Analysis \textbf{73}, 163--176
  (2014)

\bibitem[{Mai et~al.(2012)Mai, Zou and Yuan}]{Mai2012}
Mai, Q., Zou, H., Yuan, M.: A direct approach to sparse discriminant analysis
  in ultra-high dimensions.
\newblock Biometrika \textbf{99}(1), 29--42 (2012)

\bibitem[{Marks and Dunn(1974)}]{Marks1974}
Marks, S., Dunn, O.: Discriminant functions when covariance matrices are
  unequal.
\newblock Journal of the American Statistical Association \textbf{69}(346),
  555--559 (1974)

\bibitem[{Matthews(1975)}]{Matthews}
Matthews, B.W.: {Comparison of the predicted and observed secondary structure
  of T4 phage lysozyme}.
\newblock Biochimica et Biophysica Acta (BBA) - Protein Structure
  \textbf{405}(2), 442--451 (1975)

\bibitem[{Ormerod et~al.(2017)Ormerod, Stewart, Yu and Romanes}]{OrmerodCake}
Ormerod, J.T., Stewart, M., Yu, W., Romanes, S.: {Bayesian hypothesis test with
  diffused priors: Can we have our cake and eat it too?}
\newblock ArXiv  (2017)

\bibitem[{Ormerod and Wand(2010)}]{ormerodBasic}
Ormerod, J.T., Wand, M.P.: {Explaining variational approximations}.
\newblock The American Statistician \textbf{64}(2), 140--153 (2010)

\bibitem[{Perthame et~al.(2016)Perthame, Friguet and Causeur}]{Perthame2016}
Perthame, E., Friguet, C., Causeur, D.: {Stability of feature selection in
  classification issues for high-dimensional correlated data}.
\newblock Statistics and Computing \textbf{26}(4), 783--796 (2016)

\bibitem[{Perthame et~al.(2018)Perthame, Friguet and Causeur}]{FADA}
Perthame, E., Friguet, C., Causeur, D.: FADA: Variable Selection for Supervised
  Classification in High Dimension (2018).
\newblock R package version 1.3.3

\bibitem[{Safo and Ahn(2016)}]{SafoAhn}
Safo, S.E., Ahn, J.: {General sparse multi-class linear discriminant analysis}.
\newblock Computational Statistics and Data Analysis \textbf{99}, 81--90 (2016)

\bibitem[{Shaffer(1995)}]{Shaffer1995}
Shaffer, J.P.: {Multiple hypothesis testing}.
\newblock Annual Review of Psychology \textbf{46}, 561–--584 (1995)

\bibitem[{Shao et~al.(2011)Shao, Wang, Deng and Wang}]{Shao}
Shao, J., Wang, Y., Deng, X., Wang, S.: {Sparse linear discriminant analysis
  with applications to high dimensional data}.
\newblock Annals of Statistics \textbf{39}(2), 1241--1265 (2011)

\bibitem[{Singh et~al.(2002)Singh, Febbo, Ross, Jackson, Manola, Ladd, Tamayo,
  Renshaw, DAmico, Richie, Lander, Loda, Kantoff and Golub}]{Singh2002}
Singh, D., Febbo, P., Ross, K., Jackson, D., Manola, J., Ladd, C., Tamayo, P.,
  Renshaw, A., DAmico, A., Richie, J., Lander, E., Loda, M., Kantoff, P.,
  Golub, T.: Gene expression correlates of clinical prostate cancer behavior.
\newblock Cancer cell \textbf{1}(2), 203--209 (2002)

\bibitem[{Srivastava et~al.(2007)Srivastava, Gupta and
  Frigyik}]{Srivastava2007}
Srivastava, S., Gupta, M.R., Frigyik, B.A.: {Bayesian quadratic discriminant
  analysis}.
\newblock Journal of Machine Learning Research \textbf{8}(Jun), 1277--1305
  (2007)

\bibitem[{Storey(2003)}]{Storey2003}
Storey, J.D.: {The positive false discovery rate: a Bayesian interpretation and
  the q-value}.
\newblock Annals of Statistics \textbf{31}(6), 2013–--2035 (2003)

\bibitem[{Teh et~al.(2007)Teh, Newman and Welling}]{tehLDA}
Teh, Y.W., Newman, D., Welling, M.: {A collapsed variational Bayesian inference
  algorithm for latent Dirichlet allocation}.
\newblock In: Advances in Neural Information Processing Systems, vol.~19, pp.
  1353--1360. MIT Press (2007)

\bibitem[{Thomaz et~al.(2006)Thomaz, Kitani and Gillies}]{Thomaz2006}
Thomaz, C., Kitani, E., Gillies, D.: A maximum uncertainty lda-based approach
  for limited sample size problems - with applications to face recognition.
\newblock Journal of Brazilian Computer Society \textbf{12}(2), 7--18 (2006)

\bibitem[{Tibshirani et~al.(2003)Tibshirani, Hastie, Narasimhan and
  Chu}]{TibshiraniNSC}
Tibshirani, R., Hastie, T., Narasimhan, B., Chu, G.: {Class prediction by
  nearest shrunken centroids, with applications to DNA microarrays}.
\newblock Statistical Science \textbf{18}(1), 104--117 (2003)

\bibitem[{van~der Maaten and Hinton(2017)}]{vanderMaaten2017}
van~der Maaten, L., Hinton, G.: Visualising data using t-sne.
\newblock Journal of Machine Learning Research \textbf{9}, 2579--2605 (2017)

\bibitem[{Wang and Blei(2018)}]{WangBlei2018}
Wang, Y., Blei, D.: Frequentist consistency of variational bayes.
\newblock Journal of the American Statistical Association \textbf{9}, 1--15
  (2018).
\newblock \doi{10.1080/01621459.2018.1473776}

\bibitem[{Witten(2011)}]{Witten}
Witten, D.: {Classification and clustering of sequencing data using a Poisson
  model}.
\newblock Annals of Applied Statistics \textbf{5}(4), 2493--2518 (2011)

\bibitem[{Witten(2015)}]{penalizedLDA}
Witten, D.: penalizedLDA: Penalized Classification using Fisher's Linear
  Discriminant (2015).
\newblock R package version 1.1

\bibitem[{Witten and Tibshirani(2011)}]{WittenTibs}
Witten, D., Tibshirani, R.: {Penalized classification using Fisher's linear
  discriminant}.
\newblock Journal of Royal Statistical Society Series B \textbf{73}(5),
  754--772 (2011)

\bibitem[{Xu et~al.(2009)Xu, Brock and Parrish}]{Xu2009}
Xu, P., Brock, G.N., Parrish, R.S.: {Modified linear discriminant analysis
  approaches for classification of high-dimensional microarray data}.
\newblock Computational Statistics and Data Analysis \textbf{53}, 1674--1687
  (2009)

\bibitem[{Zavorka and Perrett(2014)}]{Zavorka2014}
Zavorka, S., Perrett, J.: Minimum sample size considerations for two-group
  linear and quadratic discriminant analysis with rare populations.
\newblock Communications in Statistics - Simulation and Computation
  \textbf{43}(7), 1726--1739 (2014)

\bibitem[{Zhang and Zhou(2017)}]{Zhang2017}
Zhang, A., Zhou, H.: Theoretical and computational guarantees of mean field
  variational inference for community detection.
\newblock ArXiv  (2017)

\end{thebibliography}

\end{document}